\documentclass[aps,prd,superscriptaddress,onecolumn,nofootinbib]{revtex4-2}

\usepackage{amsmath,amssymb,amsthm,mathtools}
\usepackage{graphicx,booktabs,xcolor,enumitem}
\usepackage[font=small,labelfont=bf]{caption}
\usepackage{subcaption}
\usepackage{algorithm,algpseudocode}
\makeatletter\edef\ftype@algorithm{\the\numexpr\c@float@type/2\relax}\makeatother 
\usepackage[colorlinks=true,linkcolor=blue!50!black,citecolor=blue!50!black,urlcolor=blue!50!black]{hyperref}
\usepackage{orcidlink}
\graphicspath{{./figures/}}

\usepackage{bm}
\usepackage{cleveref}

\usepackage{tikz}
\tikzset{diag/.style={baseline={([yshift=-0.6ex]current bounding box.center)}, x=0.55cm, y=0.55cm, line width=0.6pt}}
\newcommand{\dv}[2]{\fill (#1,#2) circle (2.1pt);}
\newcommand{\dl}[3]{%
    \begin{pgfinterruptboundingbox}\draw (#1,#2) .. controls +(#3-38:1.25) and +(#3+38:1.25) .. (#1,#2);\end{pgfinterruptboundingbox}%
    \path (#1,#2) ++(#3:0.55) circle (0.42);}
\newcommand{\de}[5]{\draw (#1,#2) to[bend left=#5] (#3,#4);}

\newcommand{\DFigureEight}{\tikz[diag]{\dv{0}{0}\dl{0}{0}{90}\dl{0}{0}{270}}}
\newcommand{\DSunset}{\tikz[diag]{\dv{-0.8}{0}\dv{0.8}{0}\de{-0.8}{0}{0.8}{0}{55}\de{-0.8}{0}{0.8}{0}{0}\de{-0.8}{0}{0.8}{0}{-55}}}
\newcommand{\DDumbbell}{\tikz[diag]{\dv{-0.6}{0}\dv{0.6}{0}\de{-0.6}{0}{0.6}{0}{0}\dl{-0.6}{0}{180}\dl{0.6}{0}{0}}}

\newcommand{\DFlower}{\tikz[diag]{\dv{0}{0}\dl{0}{0}{90}\dl{0}{0}{210}\dl{0}{0}{330}}}
\newcommand{\DFourFourA}{\tikz[diag]{\dv{-0.6}{0}\dv{0.6}{0}\de{-0.6}{0}{0.6}{0}{40}\de{-0.6}{0}{0.6}{0}{-40}\dl{-0.6}{0}{180}\dl{0.6}{0}{0}}}
\newcommand{\DFourFourB}{\tikz[diag]{\dv{-0.8}{0}\dv{0.8}{0}\de{-0.8}{0}{0.8}{0}{70}\de{-0.8}{0}{0.8}{0}{25}\de{-0.8}{0}{0.8}{0}{-25}\de{-0.8}{0}{0.8}{0}{-70}}}
\newcommand{\DThreeFiveA}{\tikz[diag]{\dv{-0.8}{0}\dv{0.8}{0}\de{-0.8}{0}{0.8}{0}{55}\de{-0.8}{0}{0.8}{0}{0}\de{-0.8}{0}{0.8}{0}{-55}\dl{0.8}{0}{0}}}
\newcommand{\DThreeFiveB}{\tikz[diag]{\dv{-0.6}{0}\dv{0.6}{0}\de{-0.6}{0}{0.6}{0}{0}\dl{-0.6}{0}{180}\dl{0.6}{0}{50}\dl{0.6}{0}{-50}}}
\newcommand{\DTTFa}{\tikz[diag]{\dv{-0.8}{0}\dv{0.8}{0}\dv{0}{1.2}\de{-0.8}{0}{0.8}{0}{0}\de{-0.8}{0}{0}{1.2}{25}\de{-0.8}{0}{0}{1.2}{-25}\de{0.8}{0}{0}{1.2}{25}\de{0.8}{0}{0}{1.2}{-25}}}
\newcommand{\DTTFb}{\tikz[diag]{\dv{-0.8}{0}\dv{0.8}{0}\dv{0}{1.0}\de{-0.8}{0}{0.8}{0}{30}\de{-0.8}{0}{0.8}{0}{-30}\de{-0.8}{0}{0}{1.0}{0}\de{0.8}{0}{0}{1.0}{0}\dl{0}{1.0}{90}}}
\newcommand{\DTTFc}{\tikz[diag]{\dv{-1.4}{0}\dv{0}{0}\dv{1.2}{0}\de{-1.4}{0}{0}{0}{55}\de{-1.4}{0}{0}{0}{0}\de{-1.4}{0}{0}{0}{-55}\de{0}{0}{1.2}{0}{0}\dl{1.2}{0}{0}}}
\newcommand{\DTTFd}{\tikz[diag]{\dv{-1.2}{0}\dv{0}{0}\dv{1.2}{0}\de{-1.2}{0}{0}{0}{40}\de{-1.2}{0}{0}{0}{-40}\de{0}{0}{1.2}{0}{0}\dl{-1.2}{0}{180}\dl{1.2}{0}{0}}}
\newcommand{\DTTFe}{\tikz[diag]{\dv{-1.0}{0}\dv{0}{0}\dv{1.0}{0}\de{-1.0}{0}{0}{0}{0}\de{0}{0}{1.0}{0}{0}\dl{-1.0}{0}{180}\dl{0}{0}{90}\dl{1.0}{0}{0}}}
\newcommand{\DTetra}{\tikz[diag]{\dv{0}{1.1}\dv{-0.95}{-0.55}\dv{0.95}{-0.55}\dv{0}{0}\draw (0,1.1) -- (-0.95,-0.55) -- (0.95,-0.55) -- cycle; \draw (0,0) -- (0,1.1) (0,0) -- (-0.95,-0.55) (0,0) -- (0.95,-0.55);}}
\newcommand{\DSquare}{\tikz[diag]{\dv{-0.7}{0.6}\dv{0.7}{0.6}\dv{0.7}{-0.6}\dv{-0.7}{-0.6}\de{-0.7}{0.6}{0.7}{0.6}{30}\de{-0.7}{0.6}{0.7}{0.6}{-30}\de{-0.7}{-0.6}{0.7}{-0.6}{30}\de{-0.7}{-0.6}{0.7}{-0.6}{-30}\de{0.7}{0.6}{0.7}{-0.6}{0}\de{-0.7}{0.6}{-0.7}{-0.6}{0}}}
\newcommand{\DTriTad}{\tikz[diag]{\dv{-0.7}{-0.5}\dv{0.7}{-0.5}\dv{0}{0.5}\dv{0}{1.3}\de{-0.7}{-0.5}{0.7}{-0.5}{30}\de{-0.7}{-0.5}{0.7}{-0.5}{-30}\de{-0.7}{-0.5}{0}{0.5}{0}\de{0.7}{-0.5}{0}{0.5}{0}\de{0}{0.5}{0}{1.3}{0}\dl{0}{1.3}{90}}}
\newcommand{\DDoubleTad}{\tikz[diag]{\dv{-1.3}{0}\dv{-0.5}{0}\dv{0.5}{0}\dv{1.3}{0}\de{-0.5}{0}{0.5}{0}{40}\de{-0.5}{0}{0.5}{0}{-40}\de{-1.3}{0}{-0.5}{0}{0}\de{0.5}{0}{1.3}{0}{0}\dl{-1.3}{0}{180}\dl{1.3}{0}{0}}}
\newcommand{\DStar}{\tikz[diag]{\dv{0}{0}\dv{0}{0.8}\dv{-0.69}{-0.4}\dv{0.69}{-0.4}\draw (0,0) -- (0,0.8) (0,0) -- (-0.69,-0.4) (0,0) -- (0.69,-0.4); \dl{0}{0.8}{90}\dl{-0.69}{-0.4}{210}\dl{0.69}{-0.4}{330}}}

\theoremstyle{definition}

\theoremstyle{remark}

\usepackage{placeins}

\newcommand{\dd}{\mathrm{d}}
\newcommand{\ii}{\mathrm{i}}

\begin{document}
    \title{Real-Time Path Integral Resurgence: Which Classical Path Contributes?}

    \author{Job Feldbrugge
    \orcidlink{0000-0003-2414-8707}}
    \email{jobfeldbrugge@ntu.edu.tw}
    \affiliation{Leung Center for Cosmology and Particle Astrophysics, National Taiwan University, No.~1, Sec.~4, Roosevelt Road, Taipei 10617, Taiwan}
    \affiliation{Institute for Advanced Study, 1 Einstein Drive, Princeton, NJ 08540, USA}

    \begin{abstract}
        Which complex classical paths contribute to a real-time Feynman path integral? Picard-Lefschetz theory answers this question with the intersection numbers of the thimbles, which are notoriously hard to compute in infinite dimensions. We show that resurgence provides a practical alternative. We compute the asymptotic series of the real and complex classical paths to high order, with Feynman diagrams and with a transport recursion for the cumulants, and read off the adjacency of the classical paths from the divergent tails of the series with Borel-Pad\'e methods. The South-East rule then turns adjacency into relevance. For a particle meeting a Rosen-Morse barrier, we identify the relevant complex classical paths in front of and behind the barrier, including the complex path that carries quantum tunnelling, follow ordinary and higher-order Stokes phenomena as the final point moves, and verify the predictions with the exact propagator. The resummed series reproduce the propagator with exponential accuracy. Finally, we outline how the expansion extends to relativistic path integrals governed by a Hamiltonian constraint.
    \end{abstract}

    \maketitle


    \section{Introduction}\label{sec:Introduction}
    Feynman reformulated quantum mechanics as a sum over histories \cite{Feynman:1948}. As in Young's double-slit experiment \cite{Young:1804}, the amplitude of a quantum transition follows from the interference of all the paths the system could have taken \cite{Feynman:1985}. Over the decades since, the Feynman path integral has become an invaluable tool in our exploration of the quantum universe, with applications ranging from quantum mechanics \cite{Feynman:1965} to quantum field theory \cite{Weinberg:1995,Peskin:1995} and quantum gravity \cite{Wheeler:1957,Wheeler:1961,DeWitt:1967,Hawking:1982,Hartle:1983}.

    Notwithstanding its success across physics, the path integral remains itself an object of study. In imaginary time, the Wick-rotated Euclidean path integral rests on firm ground: the Wiener measure defines it, the Feynman-Kac formula ties it to the Schr\"odinger equation \cite{Kac:1949}, and numerical methods probe it readily. The real-time path integral, which governs quantum evolution and on which we focus, enjoys no such luxury. Infinite dimensional, highly oscillatory, and at best conditionally convergent, it makes both its definition and its evaluation delicate (see the classical textbooks \cite{Feynman:1965,Schulman:1981,Kleinert:2004,Klauder:2010}). Since its inception, researchers have sought a rigorous definition, both through time discretisation \cite{Feynman:1948} and through continuum constructions \cite{DeWitt:1972,DeWitt:1974,Dewitt:1976b} (for a review of the continuum approaches, see \cite{Klauder:2003} and references therein).

    More recently, Picard-Lefschetz theory \cite{Witten:2010,Cristoforetti:2012,Fujii:2013,Feldbrugge:2017} and resurgence have given the real-time path integral a new definition, as a sum over the Lefschetz thimbles of both real and complex classical paths \cite{Dunne:2015,Dunne:2016,Feldbrugge:2023}. Recent studies show that not only the naive complex classical paths can contribute, but also their generalisations to equivalence classes, which appear when a classical path crosses a singularity of the potential \cite{Feldbrugge:2023Singular,Feldbrugge:2025RosenMorse,Feldbrugge:2026Step}. These constructions put the path integral on firmer ground, yet the real-time Feynman path integral still lacks a general, rigorous definition as an integral over paths. Even granted such a definition, one key question remains: \textit{Which classical paths are relevant to a quantum transition and need to be included, and which saddle points are irrelevant and should be ignored?}

    A clue comes from an unexpected direction: the divergence of perturbation theory. Dyson argued that the perturbative series of quantum electrodynamics cannot converge. Were it to converge for a small positive coupling, it would also converge for a small negative one, where like charges attract and the vacuum decays; the series therefore has zero radius of convergence \cite{Dyson:1952}. Bender and Wu found the same in quantum mechanics, where the perturbative coefficients of the anharmonic oscillator grow factorially \cite{Bender:1969,Bender:1973}, and Lipatov traced this growth to classical solutions in imaginary time \cite{Lipatov:1977}. The path integral inherits the structure. Every classical path carries a series in Planck's constant that dresses its classical amplitude with all quantum corrections \cite{Dewitt:1976b}, and this series diverges. Rather than a curse, the divergent tail reveals the other classical paths, and how strongly each of them couples to the first.

    The theory that decodes this information grew out of two centuries of asymptotic series. Stirling expanded the logarithm of the factorial in a series that diverges, yet whose first few terms approximate $n!$ with remarkable precision \cite{Stirling:1730}, and Euler assigned a definite value to the divergent series $\sum_n (-1)^n n!\, x^n$ \cite{Euler:1760}. Laplace turned such expansions into a method: the integral of a rapidly varying exponential concentrates near the maximum of its exponent, and an expansion around that point captures its value order by order \cite{Laplace:1812}. Stokes made asymptotic series a working tool of physics when he evaluated Airy's rainbow integral by truncating its divergent series near the smallest term \cite{Stokes:1851}. He also found that the coefficient of an exponentially small contribution jumps when a parameter crosses a line in the complex plane, the Stokes phenomenon \cite{Stokes:1864}, and Poincar\'e gave the asymptotic expansion its precise definition \cite{Poincare:1886}. A century later, Dingle observed that the late terms of a divergent series encode the early terms of the series of the neighbouring contributions \cite{Dingle:1973}, and \'Ecalle built on this observation the theory of resurgent functions \cite{Ecalle:1981}. Voros, Silverstone, and Delabaere, Dillinger, and Pham brought resurgence to the WKB method for the time-independent Schr\"odinger equation \cite{Voros:1983,Silverstone:1985,Delabaere:1997}, and Zinn-Justin and Jentschura to multi-instanton expansions \cite{ZinnJustin:2004a,ZinnJustin:2004b}. Kawai, Takei, and collaborators developed this exact WKB analysis into a theory of Stokes curves and connection formulae \cite{Kawai:2005,Aoki:2001}, which today links resurgence to exact quantisation conditions, thermodynamic Bethe ansatz equations, cluster algebras, and the spectral networks of supersymmetric gauge theories \cite{Dunne:2014,Sueishi:2020,Ito:2019,Iwaki:2014,Gaiotto:2013}. Exact WKB analysis answers the question of relevance for an ordinary differential equation: the Stokes graph, drawn from the turning points, tells which WKB solutions appear in which region. The path integral, with its infinite-dimensional space of paths, offers no such graph to draw. Berry and Howls turned resurgence into hyperasymptotics, which resums the divergent tails to exponential accuracy \cite{Berry:1990,Berry:1991,Feldbrugge:2026Wave}; Howls extended it with Delabaere to multidimensional integrals \cite{Delabaere:2002}, and with Olde Daalhuis and collaborators to differential equations and degenerate saddles \cite{OldeDaalhuis:1998,Bennett:2018}. Historically, resurgence focused on the relations among neighbouring saddle points, their \textit{adjacency}. In a recent paper, we showed that the South-East rule translates the adjacency of the saddle points of a finite-dimensional integral directly into the \textit{relevance} of these saddle points \cite{Aniceto:2026}. In the process, the South-East rule solved the problem for finite-dimension integrals. In this paper we explore the question: \textit{does this rule extend to the infinite-dimensional path integral?}

    The question of relevant complex classical paths reaches far beyond quantum mechanics. Classical paths in complex or imaginary time, the instantons, carry the effects that perturbation theory misses \cite{Coleman:1979}: the decay of a false vacuum \cite{Coleman:1977}, the tunnelling between the vacua of quantum chromodynamics \cite{Hooft:1976}, and the creation of a universe from nothing \cite{Hartle:1983,Feldbrugge:2017}. In attosecond physics, the complex trajectories of an electron in a strong laser field describe its ionisation and recollision, and the choice of saddle points decides the predicted spectra \cite{Weber:2026}. In each case, an instanton enters the calculation only when the integration contour passes through it, and a wrong choice yields a wrong answer. In quantum gravity, Kontsevich and Segal, and Witten, proposed to admit only the complex metrics on which quantum fields remain well behaved \cite{Kontsevich:2021,Witten:2021}. Such criteria rule out saddle points, but they do not decide which of the allowed saddle points contribute. Picard-Lefschetz theory does, through the intersections of the thimbles, yet these are hard to compute in infinite dimensions.

    \bigskip    

    \textit{Outline:} This paper uses resurgence to identify the relevant complex classical paths of the non-relativistic real-time path integral. \Cref{sec:PathIntegral} reviews the Feynman path integral and the Picard-Lefschetz definition. \Cref{sec:Series} develops the series of each classical path along two roads: (i) a Feynman-diagram expansion and (ii) a transport recursion method. We formulate both in terms of cumulants. The recursion delivers dozens of coefficients at a modest cost. \Cref{sec:SouthEast} applies Borel-Pad\'e methods to the classical paths of a non-relativistic particle that meets a Rosen-Morse barrier: the divergent tails reveal the adjacency of the paths, and the South-East rule turns adjacency into relevance. A spectral evaluation of the exact propagator confirms the verdict. The Borel-Pad\'e sums of the relevant paths reproduce the propagator to the accuracy of the spectral integral at large $1/\hbar$. A sweep of the final point reveals higher-order Stokes phenomena, and across the barrier, a complex path with an energy far below the barrier top carries the tunnelling amplitude. These results provide evidence, not a formal proof. Resurgence and the South-East rule rest on the analysis of finite-dimensional integrals and ordinary differential equations. For the path integral, we test them against an exactly solvable example, which supports their use but does not prove it. \Cref{sec:Relativistic} extends the recursion to Hamiltonian constraints, such as the world-line formulation of a relativistic particle and the Wheeler-DeWitt equation of quantum cosmology. \Cref{ap:RosenMorse} describes the spectral evaluation of the exact propagator.

    \section{The real-time path integral}\label{sec:PathIntegral}
    The Schr\"odinger equation governs the evolution of a non-relativistic particle of mass $m$ in a potential $V$,
    \begin{align}
        \ii \hbar \frac{\partial \psi}{\partial t} = \hat{H}\psi\,,
    \end{align}
    with the wave function $\psi(\bm{x}, t)$, the reduced Planck constant $\hbar$ and the Hamiltonian operator 
    \begin{align}
        \hat{H} = -\frac{\hbar^2}{2m} \nabla^2 + V(\bm{x})\,.
    \end{align}
    The convolution of an initial state $\psi_0$ with the Green's function 
    \begin{align}
        \psi_T(\bm{x}_1) = \int G[\bm{x}_1, \bm{x}_0; T] \psi_{0}(\bm{x}_0) \mathrm{d}\bm{x}_0\,,
    \end{align}
    solves the Schr\"odinger equation for $T > 0$, where the Green's function obeys
    \begin{align}
        \left[\ii \hbar \frac{\partial}{\partial T} - \hat{H}\right] G[\bm{x}_1, \bm{x}_0; T] = \ii \hbar\, \delta(T)\, \delta(\bm{x}_1 - \bm{x}_0)\,,
    \end{align}
    with $\hat{H}$ acting on either $\bm{x}_1$ or $\bm{x}_0$ (for a real potential, the adjoint equation in $\bm{x}_0$ takes the same form), and the limit
    \begin{align}
                \lim_{T \to 0^+}  G[\bm{x}_1, \bm{x}_0; T] =  \delta(\bm{x}_0 - \bm{x}_1)\,.
    \end{align}
    The source $\delta(T)$ selects the retarded solution, which vanishes for $T < 0$, so that the Green's function carries the factor $\Theta(T)$ with the Heaviside function $\Theta$. We leave this factor implicit in the path integral expressions below.
    
    Feynman expressed the Green's function as a sum over histories
    \begin{align}
        G[\bm{x}_1, \bm{x}_0;T] = \int_{\bm{q}(0)=\bm{x}_0}^{\bm{q}(T)=\bm{x}_1}e^{\ii S[\bm{q}(t)]/\hbar}\mathcal{D}\bm{q}(t)\,,
    \end{align}
    over all paths interpolating between $\bm{q}(0)=\bm{x}_0$ and $\bm{q}(T)=\bm{x}_1$ with the action
    \begin{align}
        S[\bm{q}(t)] = \int_0^T \left(\frac{1}{2} m \dot{\bm{q}}(t)^2 - V(\bm{q}(t))\right) \mathrm{d}t\,.
    \end{align}
    In the semiclassical limit, $\hbar \to 0$, classical mechanics arises from constructive interference at the stationary points of the action, the classical paths $\bm{q}_j$, which satisfy Newton's equation
    \begin{align}
        m \ddot{\bm{q}}(t) = - \nabla V(\bm{q}(t))\,,
    \end{align}
    and the boundary conditions $\bm{q}(0)=\bm{x}_0$ and $\bm{q}(T)=\bm{x}_1$. The solutions of the boundary value problem with a real initial velocity $\dot{\bm{q}}(0)$ form the real classical paths, central to classical mechanics \cite{Arnold:1978}. Solutions with a complex initial velocity (with non-zero imaginary parts) form the complex classical paths, which can describe classically forbidden processes such as quantum tunnelling.
    
    Following Picard-Lefschetz theory, we continuously deform the space of paths interpolating between $\bm{x}_0$ and $\bm{x}_1$ into the space of complex paths, expressing the oscillatory integral in terms of a set of convergent integrals along the descent manifolds $\mathcal{J}_j$ associated with the saddle points $\bm{q}_j$,
    \begin{align}
        G[\bm{x}_1,\bm{x}_0;T] 
        &= \sum_j n_j G^{(j)}[\bm{x}_1, \bm{x}_0; T]\,,
    \end{align}
    with the intersection numbers $n_j \in \mathbb{Z}$, a sum that includes the equivalence classes of classical paths that arise when a path crosses a singularity of the potential (see \cref{sec:SouthEast}), and the integral along the associated steepest descent manifold $\mathcal{J}_j$,
    \begin{align}
        G^{(j)}[\bm{x}_1, \bm{x}_0; T] = \int_{\mathcal{J}_j} e^{\ii S[\bm{q}(t)] / \hbar }\mathcal{D}\bm{q}(t)\,.
    \end{align}
    In this work we will assume that the critical points are non-degenerate and stay away from caustics.
    Whereas the original path integral oscillates and converged only conditionally, the integral along the descent manifold does not oscillate and decays exponentially away from the classical path (not unlike the Euclidean path integral).
    The real and complex classical paths organise the integral into a set of Wiener-like integrals. The decomposition holds exactly, not only in the limit $\hbar \to 0$, and for a real and positive Planck constant the intersection numbers $n_j$ do not depend on $\hbar$; we assume $\hbar > 0$ throughout. For a more rigorous definition of the path integral relating the path integral measure $\mathcal{D}\bm{q}(t)$ to the Brownian bridge measure $\mathrm{d}\mu_W$ (a proper $\sigma$-measure) and possible definitions of the descent manifold $\mathcal{J}_j$ in the continuum we refer to \cite{Feldbrugge:2023}.
    
    A (complex) classical path $\bm{q}_j$ is relevant when $n_j \neq 0$, and irrelevant otherwise. The non-degenerate real classical paths, central to classical mechanics, always contribute, $|n_j| = 1$. For the complex classical paths $\bm{q}_j$, capturing classically forbidden behaviour including quantum tunnelling, Picard-Lefschetz theory identifies the intersection number $n_j$ with the number of intersections of the steepest ascent manifold $\mathcal{K}_j$ of the classical path $\bm{q}_j$ with the original integration domain $C_{\bm{x}_0, \bm{x}_1}$, \textit{i.e.}, $n_j = \langle \mathcal{K}_{j}, C_{\bm{x}_0, \bm{x}_1} \rangle$. While formally elegant, the direct evaluation of the intersection number is impractical for most theories. Instead, this paper identifies the relevant complex classical paths and their intersection numbers with resurgence theory, via the South-East rule (see \cref{sec:SouthEast}).

    \section{The semiclassical expansion}\label{sec:Series}
    In the stationary-phase, or Wentzel-Kramers-Brillouin (WKB), approximation \cite{Wentzel:1926,Kramers:1926,Brillouin:1926}, we expand the action to second order around the classical paths, leading to the semiclassical approximation
    \begin{align}
        G[\bm{x}_1, \bm{x}_0; T]  \approx \Theta(T) \sum_j n_j A_j\, e^{\ii S_j/ \hbar}\,,
    \end{align}
    with the Van Vleck–Pauli–Morette amplitude \cite{VanVleck:1928,Morette:1951,Pauli:1973}
    \begin{align}
        A_j = \left(\frac{1}{2\pi\ii\hbar}\right)^{d/2} \sqrt{\det\left(-\frac{\partial^2 S_j}{\partial \bm{x}_1 \partial \bm{x}_0}\right)}\,,
    \end{align}
    where $S_j(\bm{x}_1, \bm{x}_0; T) = S[\bm{q}_j]$ denotes the classical action of the classical path as a function of its endpoints and $d$ the dimension of space. The square root follows the classical path continuously and picks up the Maslov phase $e^{-\ii\pi/2}$ at every conjugate point, counted with its multiplicity \cite{Maslov:1981}. Note that the WKB approximation inherits the intersection number from the Picard-Lefschetz decomposition. As shown by DeWitt-Morette \cite{Dewitt:1976b}, the semiclassical approximation extends to the formal expansion
    \begin{align}
        G[\bm{x}_1, \bm{x}_0; T]  \sim \Theta(T) \sum_j n_j A_j\, e^{\ii S_j/ \hbar} \sum_{m=0}^\infty a_m^{(j)} \hbar^m\,, 
        \label{eq:transseries}
    \end{align}
    with $a_0^{(j)} = 1$, in which the power series in Planck's constant dresses the classical amplitude $A_j$ with its quantum corrections. Resurgence theory customarily writes the expansion as 
    \begin{align}
        G[\bm{x}_1, \bm{x}_0; T]  \sim \Theta(T) \left(\frac{1}{2\pi\ii\hbar}\right)^{d/2} \sum_j n_j\, e^{\ii S_j/ \hbar} \sum_{m=0}^\infty T_m^{(j)} \hbar^m\,, 
    \end{align}
    with the $\hbar$-independent coefficients $T_m^{(j)} = a_m^{(j)} \sqrt{\det\left(-\partial^2 S_j/\partial \bm{x}_1 \partial \bm{x}_0\right)}$. Importantly and marked by the $\sim$, this identity is formal, as the power series is asymptotic. Its partial sums first approach the true integral, up to an optimal order near the smallest term, after which they diverge. The right-hand side of \cref{eq:transseries}, a sum over the classical paths of the exponentials $e^{\ii S_j/\hbar}$, each multiplied by its own divergent power series in $\hbar$ and weighted by its intersection number $n_j$, forms the transseries of the propagator. Even though the transseries does not converge to the propagator, it forms a unique fingerprint of the integral, from which Borel resummation or hyperasymptotic methods reconstruct the propagator \cite{Ecalle:1981,Berry:1990,Berry:1991}. We here evaluate the coefficients $a_m^{(j)}$ along two distinct roads, a Feynman-diagram expansion (\cref{sec:Diagrams}) and a transport recursion (\cref{sec:Recursion}).

    \subsection{The Feynman diagrams}\label{sec:Diagrams}
    We first review DeWitt-Morette's semiclassical expansion \cite{Dewitt:1976b} whose core can be traced back to Feynman and Hibbs's textbook \cite{Feynman:1965}. Writing every path as the classical path and a fluctuation $\bm{q}(t) = \bm{q}_j(t) + \sqrt{\hbar} \bm{v}(t)$ with the boundary condition $\bm{v}(0) = \bm{v}(T) = \bm{0}$, the exponent of the integrand assumes the form 
    \begin{align}
        \frac{\ii}{\hbar} S[\bm{q}(t)] = 
        \frac{\ii}{\hbar} S_j + 
        \frac{\ii}{2} \int_0^T \left(m \dot{\bm{v}}(t)^2 - \bm{v}(t)^{\mathsf T} [\mathcal{H} V(\bm{q}_j(t))] \bm{v}(t)\right)\mathrm{d}t 
        - \ii \sum_{n  = 3}^\infty \frac{\hbar^{\frac{n}{2} - 1}}{n!} \int_0^T V_{a_1\cdots a_n}(\bm{q}_j(t)) v_{a_1}(t) \cdots v_{a_n}(t) \mathrm{d}t
    \end{align}
    with the Hessian of the potential $\mathcal{H} V$ and the $n$-th order derivative of the potential $V_{a_1 \cdots a_n} = \partial_{a_1} \cdots \partial_{a_n} V$ where $a_1,\dots,a_n \in \{1,\dots, d\}$. Here and below, a repeated index is summed over $\{1, \dots, d\}$ (Einstein's summation convention). Upon expanding the exponential function of the cubic and higher-order terms, the contribution $G^{(j)}$ of the thimble of $\bm{q}_j$ assumes the form 
    \begin{align}
        G^{(j)}[\bm{x}_1, \bm{x}_0; T] = &\frac{1}{\hbar^{d/2}}e^{\ii S_j / \hbar} 
        \int_{\bm{v}(0)=\bm{0}}^{\bm{v}(T)=\bm{0}} e^{
             \frac{\ii}{2} \int_0^T \left(m \dot{\bm{v}}(t)^2 - \bm{v}(t)^{\mathsf T} [\mathcal{H} V(\bm{q}_j(t))] \bm{v}(t)\right)\mathrm{d}t} \nonumber\\
             &\times \sum_{r=0}^\infty \frac{1}{r!}\left[
                    - \ii \sum_{n = 3}^\infty \frac{\hbar^{\frac{n}{2} - 1}}{n!} \int_0^T V_{a_1\cdots a_n}(\bm{q}_j(t)) v_{a_1}(t) \cdots v_{a_n}(t) \mathrm{d}t
             \right]^r \mathcal{D}\bm{v}(t)\,.
    \end{align}
    The $r=0$ term reduces to the Gaussian integral of a particle in a time-dependent harmonic oscillator, 
    \begin{align}
        \frac{1}{\hbar^{d/2}} \int_{\bm{v}(0)=\bm{0}}^{\bm{v}(T)=\bm{0}} e^{
             \frac{\ii}{2} \int_0^T \left(m \dot{\bm{v}}(t)^2 - \bm{v}(t)^{\mathsf T} [\mathcal{H} V(\bm{q}_j(t))] \bm{v}(t)\right)\mathrm{d}t} \mathcal{D}\bm{v}(t) 
    \end{align}
    which, for a suitably normalised measure $\mathcal{D}\bm{v}$, coincides with the Van Vleck–Pauli–Morette amplitude $A_j$ and recovers the WKB approximation. The higher-order corrections, corresponding to integer powers of Planck's constant, reduce to integrals over the moments 
    \begin{align}
        \langle v_{a_1}(t_1) \cdots v_{a_n}(t_n) \rangle = \frac{\int_{\bm{v}(0)=\bm{0}}^{\bm{v}(T)=\bm{0}} e^{
             \frac{\ii}{2} \int_0^T \left(m \dot{\bm{v}}(t)^2 - \bm{v}(t)^{\mathsf T} [\mathcal{H} V(\bm{q}_j(t))] \bm{v}(t)\right)\mathrm{d}t}\, v_{a_1}(t_1) \cdots v_{a_n}(t_n)\, \mathcal{D}\bm{v}(t)}{\int_{\bm{v}(0)=\bm{0}}^{\bm{v}(T)=\bm{0}} e^{
             \frac{\ii}{2} \int_0^T \left(m \dot{\bm{v}}(t)^2 - \bm{v}(t)^{\mathsf T} [\mathcal{H} V(\bm{q}_j(t))] \bm{v}(t)\right)\mathrm{d}t}\, \mathcal{D}\bm{v}(t)}\,.
    \end{align}
    
\begin{figure}
        \centering
        \includegraphics{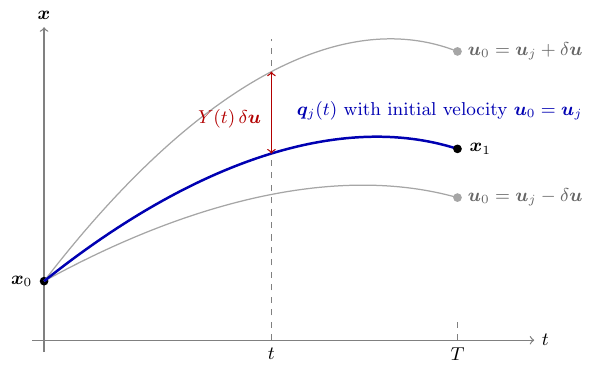}
        \caption{The trajectories from $x_0$ form a family, and the initial velocity $u_0$ labels its members. The classical path $q_j$ (thick) reaches $x_1$ at the time $T$. At the time $t$, the neighbour with the initial velocity $u_0 = u_j + \delta u$ lies a distance $y(t)\,\delta u$ away, with $y = \partial q/\partial u_0$ the one-dimensional Jacobi field. The transport recursion (\cref{sec:Recursion}) uses $u_0$ to label the points near the classical path.}
        \label{fig:lagrangian}
    \end{figure}

    Wick's theorem reduces every Gaussian moment to products of the two-point function $\langle v_a(t_1) v_b(t_2) \rangle = - \ii \mathcal{G}^{(j)}_{ab}(t_1,t_2)$. For example, with the shorthand $V_{a_1\cdots a_n}(t) = V_{a_1\cdots a_n}(\bm{q}_j(t))$, the first correction assumes the form 
    \begin{align}
        a_1^{(j)} &= \frac{\ii}{8}\int_0^T V_{abcd}(t)\, \mathcal{G}^{(j)}_{ab}(t,t)\, \mathcal{G}^{(j)}_{cd}(t,t)\, \dd t
        \nonumber\\
        &\phantom{=}- \frac{\ii}{12}\int_0^T \int_0^T V_{abc}(t_1)\, V_{def}(t_2)\, \mathcal{G}^{(j)}_{ad}(t_1,t_2)\, \mathcal{G}^{(j)}_{be}(t_1,t_2)\, \mathcal{G}^{(j)}_{cf}(t_1,t_2)\, \dd t_1\, \dd t_2\nonumber\\
        &\phantom{=}- \frac{\ii}{8}\int_0^T \int_0^T V_{abc}(t_1)\, \mathcal{G}^{(j)}_{ab}(t_1,t_1)\, \mathcal{G}^{(j)}_{cd}(t_1,t_2)\, V_{def}(t_2)\, \mathcal{G}^{(j)}_{ef}(t_2,t_2)\, \dd t_1\, \dd t_2\,,
        \label{eq:sigma1diagram}
    \end{align}
    where $\mathcal{G}^{(j)}(t_1,t_2)$ is the Green's function of the Jacobi operator,
    \begin{align}
        m\,\partial_{t_1}^2 \mathcal{G}^{(j)}(t_1,t_2) + \left[\mathcal{H} V(\bm{q}_j(t_1))\right] \mathcal{G}^{(j)}(t_1,t_2) = \mathbb{I}\,\delta(t_1-t_2)\,,\qquad \mathcal{G}^{(j)}(0,t_2) = \mathcal{G}^{(j)}(T,t_2) = 0\,,
        \label{eq:green}
    \end{align}
    for $0 < t_2 < T$ with the $d\times d$ identity matrix $\mathbb{I}$ (for a derivation see \cite{Feynman:1965}). The conditions in $t_2$ follow from the symmetry $\mathcal{G}^{(j)}(t_1,t_2) = \mathcal{G}^{(j)}(t_2,t_1)^{\mathsf T}$. The Jacobi equation lies at the heart of classical mechanics: it describes how a congruence of classical paths evolves in time (\cref{fig:lagrangian}). Write the classical path from $\bm{x}_0$ with the initial velocity $\bm{u}_0$ as $\bm{q}(\bm{x}_0,\bm{u}_0,t)$, and let $\bm{u}_j$ denote the initial velocity of the classical path $\bm{q}_j$, so that $\bm{q}_j(t) = \bm{q}(\bm{x}_0, \bm{u}_j, t)$. The derivative with respect to the initial velocity, $Y(t) = \tfrac{\partial \bm{q}(\bm{x}_0, \bm{u}_0, t)}{\partial \bm{u}_0}$ at $\bm{u}_0 = \bm{u}_j$, satisfies the Jacobi equation 
    \begin{align}
        m \ddot{Y} (t) + \left[\mathcal{H} V(\bm{q}_j(t))\right] Y(t) = 0\,. \label{eq:Y}
    \end{align}
    The variation of the initial velocity around $\bm{u}_j$ yields the solution $Y_1 = Y$ with the boundary conditions $Y_1(0)=0$ and $\dot{Y}_1(0) = \mathbb{I}$. Likewise, the reversed classical path, which starts at $\bm{x}_1$ at the time $T$ and reaches $\bm{x}_0$ at $t=0$, varied with respect to its velocity at $t=T$, yields the solution $Y_2$ with the boundary conditions $Y_2(T)=0$ and $\dot{Y}_2(T) = - \mathbb{I}$. Together, the two solutions build the Green's function by the Wronskian method
    \begin{align}
        \mathcal{G}^{(j)}(t_1, t_2) = \begin{cases}
            Y_1(t_1)\,(mW)^{-\mathsf{T}}\, Y_2(t_2)^{\mathsf T}\,, & t_1 \leq t_2\,,\\
            Y_2(t_1)\,(mW)^{-1}\, Y_1(t_2)^{\mathsf T}\,, & t_1 > t_2\,,
        \end{cases}
        \label{eq:greenY}
    \end{align}
    where $M^{-\mathsf T}$ denotes the inverse of the transpose of the matrix $M$, and the Wronskian
    \begin{align}
        W = Y_1^{\mathsf T}\dot{Y}_2 - \dot{Y}_1^{\mathsf T} Y_2
    \end{align}
    is independent of time. The classical paths do not only organise the path integral in the Picard-Lefschetz formula, but also govern the quantum corrections $a_m^{(j)}$ to the functional determinant.

\begin{figure}
        \centering
        \includegraphics{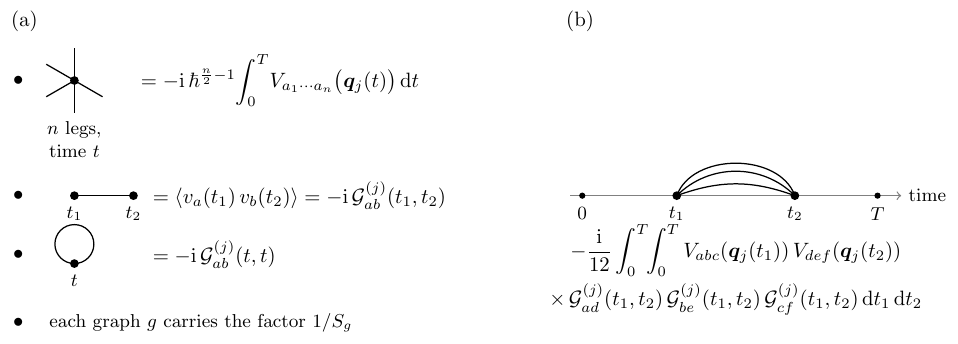}
        \caption{(a) The Feynman rules of the expansion around a classical path: a vertex with $n$ legs at the time $t$, a line, a loop, and the symmetry factor of the graph. (b) A diagram lives on the time interval $[0, T]$. The sunset diagram connects two cubic insertions at the times $t_1$ and $t_2$ along the classical path $\bm{q}_j$ with three Green's functions, and contributes the value shown, which includes the weight $(-\ii)^{V_g+E_g}/S_g = -\ii/12$.}
        \label{fig:rules}
    \end{figure}

    Expanding $G^{(j)}$ in powers of $\hbar$ yields a Feynman-diagram expansion with the rules:
    \begin{itemize}
        \item a \emph{vertex} with $n \geq 3$ legs stands for the insertion $-\ii\,\hbar^{\frac{n}{2}-1}V_{a_1\cdots a_n}(\bm{q}_j(t))$ at the time $t$, integrated over $[0, T]$;
        \item a \emph{line} from the leg $a$ at the time $t_1$ to the leg $b$ at the time $t_2$ stands for $-\ii\, \mathcal{G}^{(j)}_{ab}(t_1,t_2)$, and a \emph{loop} from a vertex to itself for $-\ii\, \mathcal{G}^{(j)}_{ab}(t,t)$;
        \item the graph carries the factor $1/S_g$, with its \emph{symmetry factor} $S_g$, the number of permutations of vertices and lines that leave the graph unchanged. The factors $1/n!$ of the vertices and $1/r!$ of the expansion combine with the number of Wick contractions into this factor.
    \end{itemize}
    The indices of the legs contract along the lines. For an illustration see \cref{fig:rules}.

    Finally, by the linked-cluster theorem, the logarithm of the sum of all vacuum graphs equals the sum of the connected ones \cite{Goldstone:1957,Kubo:1962,Peskin:1995}. Writing the expansion 
    \begin{align}
        \sum_{m=0}^\infty a_m^{(j)} \hbar^m = e^{\sum_{m=1}^\infty \sigma_m^{(j)} \hbar^m}
    \end{align}
    in terms of the cumulants $\sigma_m^{(j)}$, we obtain the identification
    \begin{align}
        \sigma_m^{(j)} = \sum_{\substack{\text{connected } g\\ E_g - V_g = m}} \frac{(-\ii)^{V_g + E_g}}{S_g} \int_0^T\!\!\cdots\!\int_0^T \prod_{v \in g} V_{a_{v,1}\cdots a_{v,n_v}}\big(\bm{q}_j(t_v)\big)\, \dd t_v \prod_{\ell \in g} \mathcal{G}^{(j)}_{a_{\ell} b_{\ell}}(t_{v_\ell}, t_{w_\ell})\,,
        \label{eq:sigmadiagrams}
    \end{align}
    where the sum runs over the connected vacuum graphs $g$ with $V_g$ vertices and $E_g$ lines, the vertex $v$ sits at the time $t_v$ and has $n_v \geq 3$ legs with the indices $a_{v,1}, \dots, a_{v,n_v}$, and the line $\ell$ joins a leg with the index $a_\ell$ at the vertex $v_\ell$ to a leg with the index $b_\ell$ at the vertex $w_\ell$, with $v_\ell = w_\ell$ for a loop. Every index appears twice, once at a vertex and once at a line, and is therefore summed. A graph with $E_g - V_g = m$ has $m + 1$ loops and carries the power $\hbar^m$, as $\sum_v (n_v/2 - 1) = E_g - V_g$ when every leg ends on a line. Each graph becomes a picture of the quantum correction: insertions of the potential's derivatives at a few moments along the classical path, tied together by the Green's function. The diagrams with two loops make up the first-order cumulant and summarise \cref{eq:sigma1diagram}:
    \begin{align}
        \sigma_1^{(j)} = \frac{\ii}{8}\; \DFigureEight \;-\; \frac{\ii}{12}\; \DSunset \;-\; \frac{\ii}{8}\; \DDumbbell\,.
    \end{align}
    DeWitt-Morette \cite{Dewitt:1976b} and Gaspard and coworkers \cite{Gaspard:1993,Gaspard:1995} derive the same result. At order $\hbar^2$, fifteen connected three-loop graphs contribute, with the weights $(-\ii)^{V_g+E_g}/S_g$ of \cref{eq:sigmadiagrams},
    \begin{align}
        \sigma_2^{(j)} = \;&\frac{1}{48}\, \DFlower \;-\; \frac{1}{16}\, \DFourFourA \;-\; \frac{1}{48}\, \DFourFourB \;-\; \frac{1}{12}\, \DThreeFiveA \;-\; \frac{1}{16}\, \DThreeFiveB \nonumber\\[1mm]
        &+\; \frac{1}{8}\, \DTTFa \;+\; \frac{1}{8}\, \DTTFb \;+\; \frac{1}{12}\, \DTTFc \;+\; \frac{1}{8}\, \DTTFd \;+\; \frac{1}{16}\, \DTTFe \nonumber\\[1mm]
        &-\; \frac{1}{24}\, \DTetra \;-\; \frac{1}{16}\, \DSquare \;-\; \frac{1}{8}\, \DTriTad \;-\; \frac{1}{16}\, \DDoubleTad \;-\; \frac{1}{48}\, \DStar\,.
        \label{eq:sigma2diagram}
    \end{align}
    They combine one sextic vertex; two quartic vertices; a cubic and a quintic vertex; two cubic and a quartic vertex; or four cubic vertices. Each graph integrates over up to four insertion times. The number or required Feynman diagrams grows factorially with $m$.

    \subsection{The transport recursion}\label{sec:Recursion}
    The diagrammatic expansion beautifully formulates the asymptotic series of a saddle point $\bm{q}_j$ in terms of its Green's function $\mathcal{G}^{(j)}(t_1,t_2)$, but is expensive as the number of diagrams grows factorially with the order, each corresponding to the numerical evaluation of a multidimensional integral. Following Maslov and Fedoriuk \cite{Maslov:1981}, we leverage the Schr\"odinger equation to evaluate the asymptotic series of a classical path in terms of a recursion relation. Whereas Maslov and Fedoriuk expressed their recursion relations in terms of the coefficients $a_m^{(j)}$, we here work with the cumulants $\sigma_m^{(j)}$ associated with the connected diagrams. This method is less intuitive but circumvents the factorial growth of terms present in the Feynman expansion. The same distinction can be seen when computing asymptotic series of finite-dimensional integrals using Gaussian moments and the associated differential equations \cite{Crew:2026}.
    
    The derivation of the Schr\"odinger equation from the path integral holds for every integration contour that does not change under small changes of the endpoint $\bm{x}_1$ and the time $t$. The thimble $\mathcal{J}_j$ forms such a contour, so each thimble integral solves the Schr\"odinger equation on its own,
    \begin{align}
        \ii \hbar \frac{\partial G^{(j)}[\bm{x}_1,\bm{x}_0; t]}{\partial t} = -\frac{\hbar^2}{2m} \nabla_{\bm{x}_1}^2 G^{(j)}[\bm{x}_1,\bm{x}_0;t] + V(\bm{x}_1)G^{(j)}[\bm{x}_1,\bm{x}_0;t]\,.
    \end{align}
    To solve it, write the thimble integral as an exponential, $G^{(j)}[\bm{x}_1, \bm{x}_0;t] = A_j e^{\ii S_j/\hbar + L^{(j)}} = e^{\Phi}$, with the exponent $\Phi = \ii S_j/\hbar + \alpha^{(j)} + L^{(j)}$, the logarithm of the amplitude $\alpha^{(j)} = \ln A_j$, and the remainder $L^{(j)} = \sum_{m=1}^\infty \sigma_m^{(j)} \hbar^m$. The Laplacian of an exponential, $\nabla_{\bm{x}_1}^2 e^{\Phi} = e^{\Phi}\left[\nabla_{\bm{x}_1}^2\Phi + (\nabla_{\bm{x}_1}\Phi)^2\right]$, turns the linear Schr\"odinger equation for $G^{(j)}$ into a non-linear equation for its logarithm,
    \begin{align}
        \ii\hbar\, \partial_t \Phi = -\frac{\hbar^2}{2m}\left[\nabla_{\bm{x}_1}^2\Phi + (\nabla_{\bm{x}_1}\Phi)^2\right] + V\,.
    \end{align}
    Inserting the three parts of $\Phi$ yields
    \begin{align}
        -\partial_t S_j + \ii\hbar\,\partial_t(\alpha^{(j)} + L^{(j)}) = \frac{(\nabla_{\bm{x}_1} S_j)^2}{2m} + V &- \frac{\ii\hbar}{2m}\left[\nabla_{\bm{x}_1}^2 S_j + 2\,\nabla_{\bm{x}_1} S_j\cdot\nabla_{\bm{x}_1}(\alpha^{(j)} + L^{(j)})\right] \nonumber\\
        &- \frac{\hbar^2}{2m}\left[\nabla_{\bm{x}_1}^2(\alpha^{(j)} + L^{(j)}) + \big(\nabla_{\bm{x}_1}(\alpha^{(j)} + L^{(j)})\big)^2\right]\,.
    \end{align}
    The action $S_j$ does not depend on $\hbar$, and neither does the amplitude $\alpha^{(j)}$, apart from the constant $-\frac{d}{2}\ln(2\pi\ii\hbar)$ that drops out of every derivative, while the remainder $L^{(j)}$ starts at the order $\hbar$. Sorting by powers of $\hbar$, we find
    \begin{itemize}
        \item at order $\hbar^0$, the action obeys the Hamilton-Jacobi equation, $\partial_t S_j + (\nabla_{\bm{x}_1} S_j)^2/2m + V = 0$;
        \item at order $\hbar^1$, the amplitude obeys the continuity equation, $\frac{\mathrm{D}\alpha^{(j)}}{\dd t} = -\nabla_{\bm{x}_1}^2 S_j/2m$;
        \item the remainder obeys
        \begin{align}
            \frac{\mathrm{D} L^{(j)}}{\dd t} = \frac{\ii\hbar}{2m}\left[\nabla_{\bm{x}_1}^2 \alpha^{(j)} + (\nabla_{\bm{x}_1} \alpha^{(j)})^2 + 2\,\nabla_{\bm{x}_1}\alpha^{(j)}\cdot\nabla_{\bm{x}_1} L^{(j)} + \nabla_{\bm{x}_1}^2 L^{(j)} + (\nabla_{\bm{x}_1} L^{(j)})^2\right]\,,
        \end{align}
    \end{itemize}
    with the co-moving derivative $\frac{\mathrm{D}}{\dd t} = \partial_t + \dot{\bm{x}}\cdot\nabla_{\bm{x}_1}$, which differentiates along the classical paths with the velocity $\dot{\bm{x}} = \nabla_{\bm{x}_1} S_j/m$. The expansion of $L^{(j)}$ in $\hbar$ yields the transport recursion
    \begin{align}
        \frac{\mathrm{D} \sigma_1^{(j)}}{\dd t} &= \frac{\ii}{2m}\left[\nabla_{\bm{x}_1}^2 \alpha^{(j)} + (\nabla_{\bm{x}_1}\alpha^{(j)})^2\right],\nonumber\\
        \frac{\mathrm{D} \sigma_{m+1}^{(j)}}{\dd t} &= \frac{\ii}{2m}\left[\nabla_{\bm{x}_1}^2 \sigma_m^{(j)} + 2\,\nabla_{\bm{x}_1}\alpha^{(j)}\cdot\nabla_{\bm{x}_1} \sigma_m^{(j)} + \sum_{k=1}^{m-1} \nabla_{\bm{x}_1}\sigma_k^{(j)}\cdot\nabla_{\bm{x}_1} \sigma_{m-k}^{(j)}\right].
        \label{eq:transport}
    \end{align}
    As the propagator approaches the free propagator in the limit $t \to 0^+$, we obtain the boundary conditions $\sigma_m^{(j)} = 0$ completing the system. The same substitution turns the time-independent Schr\"odinger equation in one dimension into the Riccati equation of the WKB method, whose formal solution exact WKB analysis Borel resums \cite{Voros:1983,Kawai:2005}. The transport recursion forms its time-dependent counterpart. It expands the propagator around a classical path rather than a stationary wave around one branch of the momentum $\pm\sqrt{2m(E - V)}$, and the free propagator at $t \to 0^+$ fixes the integration constants that exact WKB analysis fixes with a reference point, often a turning point.
    
    The recursion requires the gradients of $\alpha^{(j)}$ and $\sigma_m^{(j)}$ at the classical path $\bm{q}_j$. We efficiently obtain these variations using the Jacobian $Y(t)$. Label the point $\bm{x}_1 = \bm{q}(\bm{x}_0, \bm{u}_0, t)$ by the time $t$ and the initial velocity $\bm{u}_0$ of the classical path from $\bm{x}_0$ that reaches it: the classical path $\bm{q}_j$ carries the label $\bm{u}_j$, and its neighbours the labels $\bm{u}_0 = \bm{u}_j + \delta\bm{u}$ (\cref{fig:lagrangian}). In these Lagrangian coordinates, the co-moving derivative reduces to the ordinary time derivative at fixed $\bm{u}_0$, as the velocity field $\nabla_{\bm{x}_1} S_j/m$ carries every point along its own path, and the chain rule converts the spatial gradient into a gradient in velocity space,
    \begin{align}
        \nabla_{\bm{x}_1} = Y(t)^{-\mathsf T}\, \nabla_{\bm{u}_0}\,,
    \end{align}
    with the Jacobi matrix of \cref{eq:Y}. The continuity equation then integrates in closed form. The Laplacian of the action measures the divergence of the velocity field, $\nabla_{\bm{x}_1}^2 S_j/m = \nabla_{\bm{x}_1} \cdot \dot{\bm{x}} = \frac{\dd}{\dd t} \ln\det Y$, so that
    \begin{align}
        \alpha^{(j)}(\bm{x}_0, \bm{u}_0, t) = -\frac{1}{2} \ln \det Y(\bm{x}_0, \bm{u}_0, t) + \text{const}\,,
    \end{align}
    where the constant drops out of every gradient. The transport recursion now turns into ordinary integrals over time, one for every label $\bm{u}_0$, and the cumulants of the classical path follow at the label $\bm{u}_0 = \bm{u}_j$ and the time $t = T$. The first two cumulants read
    \begin{align}
        \sigma_1^{(j)}(\bm{x}_0, \bm{u}_0, t) &= \frac{\ii}{2m} \int_0^t \left[\nabla_{\bm{x}_1}^2 \alpha^{(j)} + \big(\nabla_{\bm{x}_1} \alpha^{(j)}\big)^2\right](\bm{x}_0, \bm{u}_0, t')\, \dd t'\,,\\
        \sigma_2^{(j)}(\bm{x}_0, \bm{u}_0, t) &= \frac{\ii}{2m} \int_0^t \left[\nabla_{\bm{x}_1}^2 \sigma_1^{(j)} + 2\, \nabla_{\bm{x}_1}\alpha^{(j)} \cdot \nabla_{\bm{x}_1} \sigma_1^{(j)}\right](\bm{x}_0, \bm{u}_0, t')\, \dd t'\,.
        \label{eq:sigma12transport}
    \end{align}
    The second cumulant reveals the structure of the recursion: its source contains the gradients of $\sigma_1^{(j)}$, so $\sigma_2^{(j)}$ needs the first cumulant not only along $\bm{q}_j$ but along its neighbours. The higher cumulants need ever higher derivatives with respect to $\bm{u}_0$. To supply them, carry every quantity as a truncated power series in $\delta\bm{u} = \bm{u}_0 - \bm{u}_j$: the trajectory $\bm{q}(\bm{x}_0, \bm{u}_0, t)$, its velocity, and the cumulants $\sigma_m^{(j)}(\bm{x}_0, \bm{u}_0, t)$. Newton's equation, composed with the series, advances the trajectory, and its coefficients deliver the Jacobi matrix $Y$ together with its derivatives $\partial_{\bm{u}_0} Y, \partial_{\bm{u}_0}^2 Y, \dots$. The recursion \eqref{eq:transport}, with $\nabla_{\bm{x}_1} = Y^{-\mathsf T}\nabla_{\bm{u}_0}$ acting on the series, advances the cumulants. The constant terms of the series describe the classical path $\bm{q}_j$, the higher terms its neighbours. A simple count fixes the length of the series. To compute the cumulants up to the order $M$, $\sigma_1^{(j)}, \dots, \sigma_M^{(j)}$, note that every order of the recursion applies two derivatives, and only the constant term of $\sigma_M^{(j)}$ survives at the end. Hence $\sigma_m^{(j)}$ needs the degree $2(M - m)$ in $\delta\bm{u}$, the amplitude $\alpha^{(j)}$ the degree $2M$, and the trajectory the degree $2M + 1$. The work grows polynomially with the order $M$, rather than factorially as for the diagrams.

    The neighbouring paths focus where $\det Y$ vanishes, and at such a caustic, or conjugate point, the conversion $\nabla_{\bm{x}_1} = Y^{-\mathsf T}\nabla_{\bm{u}_0}$ breaks down. The amplitude $\alpha^{(j)} = -\frac{1}{2}\ln\det Y$ diverges logarithmically, and the sources of the recursion, built from powers of $Y^{-1}$, diverge as well. Yet the Feynman diagrams remain finite there \cite{Dewitt:1976b}, as the Green's function stays finite as long as $\det Y(T) \neq 0$, and so do the cumulants at the time $T$. For a real path, the caustics lie on the real time axis between $0$ and $T$. For a complex path, they move into the complex plane, but they often lie close to the real axis, where the large entries of $Y^{-1}$ cost accuracy through cancellations. The cure lies in the analyticity of the equations. For an analytic potential, the trajectory $\bm{q}(\bm{x}_0, \bm{u}_0, t)$, the Jacobi matrix $Y$, the amplitude $\alpha^{(j)}$, and the cumulants $\sigma_m^{(j)}(\bm{x}_0, \bm{u}_0, t)$ continue analytically into complex time, and Newton's equation, the Jacobi equation, and the recursion \eqref{eq:transport} hold along any path in the complex $t$-plane. To avoid the caustics, deform the time domain $[0, T]$ into a contour that leaves the real axis before each caustic and rejoins it after, and integrate all equations along this contour. At a simple zero of $\det Y$, the cumulants have poles but no branch points, so their values at the time $T$ do not depend on the side of the detour. The side matters only for the amplitude: Feynman's prescription $T \to T(1 - \ii\epsilon)$ selects the detour below the real axis, along which $\alpha^{(j)} = -\frac{1}{2}\ln\det Y$ acquires the Maslov phase $-\ii\pi/2$ at every caustic. The same deformation steers a complex path around the singularities of the potential \cite{Feldbrugge:2026Step}.

    Finally note that the recursion relation provides a means to compute the transseries of the real-time path integral without reference to its definition as an integral over paths. The Hamilton-Jacobi equation provides the classical paths as its characteristics. The Van Vleck-Pauli-Morette amplitude follows from the first-order term in $\hbar$. The quantum corrections in the form of the cumulants follow from the higher-order terms with the limit free particle limit $\hbar \to 0$. What is missing are the intersection numbers $n_j$, for the path integral along any set of Lefschetz thimbles solves the Schr\"odinger equation.

    \section{Which classical path contributes?}\label{sec:SouthEast}
    The semiclassical expansion reveals the adjacency relations of the classical paths, and the South-East rule turns them into relevance. We illustrate the discussion with the path integral of a non-relativistic particle in a one-dimensional Rosen-Morse barrier \cite{Rosen:1932} (sometimes known as the P\"oschl-Teller potential \cite{Poschl:1933})
    \begin{align}
        V(x) = \frac{1}{\cosh^2 x}\,.
    \end{align}
    The Rosen-Morse barrier makes an ideal test: its real-time path integral includes an infinite tower of closed-form classical paths, yet spectral methods solve it exactly (see \cref{ap:RosenMorse}). For a recent and detailed exposition of Rosen-Morse theory and its complex classical paths, we refer to \cite{Feldbrugge:2025RosenMorse} and references therein.

    Newton's equation
    \begin{align}
        \ddot{q}(t) = \frac{2\sinh q(t)}{\cosh^3 q(t)}\,,
    \end{align}
    with the initial values $q(0)=x_0$ and $\dot{q}(0)=u_0$, and the unit mass $m=1$, linearises under the change of variables $s(t) = \sinh q(t)$ into the equation of motion, $\ddot{s}(t) = 2E\, s(t)$, leading to the classical path
    \begin{align}
        \sinh q(t) = c_1 \sqrt{\frac{E - 1}{E}}\, \sinh\left[\sqrt{2E}\,(t - t_c)\right]\,,
    \end{align}
    with the sign $c_1 = \pm 1$ and the energy
    \begin{align}
        E = \frac{u_0^2}{2} + \frac{1}{\cosh^2 x_0}\,.
    \end{align}
    The initial position fixes the time
    \begin{align}
        t_c =- \frac{c_1}{\sqrt{2E}}\, \sinh^{-1}\left[\sqrt{\frac{E}{E - 1}}\, \sinh x_0\right]\,,
    \end{align}
    while the initial velocity fixes $c_1$ through $\dot{q}(0) = u_0$. 

    The classical paths that reach $q(T) = x_1$, found by solving for the initial velocity $u_0$, govern the semiclassical expansion of the path integral. The Rosen-Morse path integral generally has one or three real classical paths and an infinite tower of complex classical paths, respectively characterised by real and complex initial velocities. As $x_0$ and $x_1$ vary smoothly, the complex classical paths may encounter a pole of the analytic continuation of the potential on the imaginary axis at $x = \ii \pi \left(n + \frac{1}{2}\right)$ for integer $n$. The path meets a pole at the times $t_*$ where $\sinh q(t_*) = \pm \ii$. Each such point $t_*$ forms a branch point of the classical path $q(t)$ and a pole of the Lagrangian in the complex $t$-plane. A singularity crossing occurs when $t_*$ crosses the time contour $[0, T]$.

    \begin{figure}
        \centering
        \includegraphics{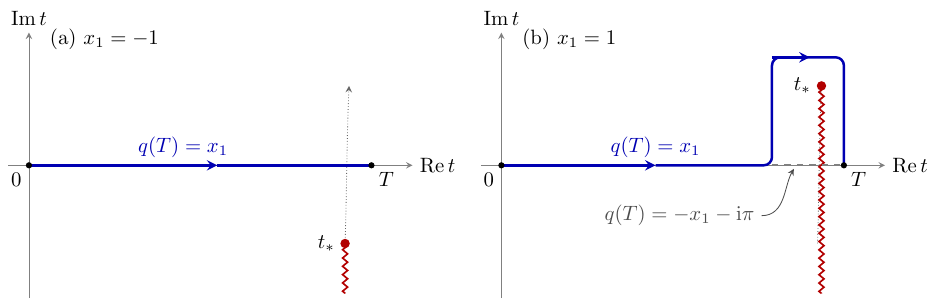}
        \caption{The time contour of the complex classical path $q_1$ of the Rosen-Morse barrier ($x_0 = -5$, $T = 10$) in the complex time plane. (a) At $x_1 = -1$, the branch point $t_*$ of $q(t)$, where the path reaches a pole of the potential and the Lagrangian has a pole, lies below the real time axis, and the path along the real time axis (blue) reaches $x_1$. The dotted curve traces $t_*$ as $x_1$ increases from $-1$ to $1$. (b) At $x_1 = 1$, beyond the singularity crossing, $t_*$ lies above the real time axis. The path along the real time axis (dashed) crosses the branch cut (red) and ends at $-x_1 - \ii\pi$. The deformed contour (blue) keeps $t_*$ on its original side, reaches $q(T) = x_1$, and gives the analytically continued action $S = 0.0601 + 2.8224\,\ii$. The two actions differ by $2\pi\ii$ times the residue of the Lagrangian at $t_*$, $\Delta S = \ii\sqrt{2}\,\pi$.}
        \label{fig:timecontour}
    \end{figure}

    For $x_0 = -5$ and $T = 10$, the relevant complex classical path $q_1$ makes the effect visible (see \cref{fig:timecontour}). At $x_1 = -1$, its branch point sits below the real time axis, at $t_* = 9.230 - 0.854\,\ii$, and the path along the real axis reaches $x_1$. As the final point moves to the right, the branch point climbs, crosses the real time axis at $x_1 \approx -0.025$, and reaches $t_* = 9.345 + 0.868\,\ii$ at $x_1 = 1$. Beyond this singularity crossing, the path along the real time axis no longer reaches $x_1$. Instead, it ends at $-x_1 - \ii\pi$, with the action $0.0601 - 1.6205\,\ii$. Yet the complex path still contributes. Deform the time contour such that it keeps $t_*$ on its original side, and the path reaches $q(T) = x_1$ again, with the action $S_1 = 0.0601 + 2.8224\,\ii$, the analytic continuation of the action from $x_1 = -1$. The saddle point survives the crossing as an equivalence class of classical paths along deformed time contours \cite{Feldbrugge:2023Singular,Feldbrugge:2023,Feldbrugge:2026Step}.

    \subsection{Adjacency relations}
\begin{figure}
        \centering
        \includegraphics[width=\textwidth]{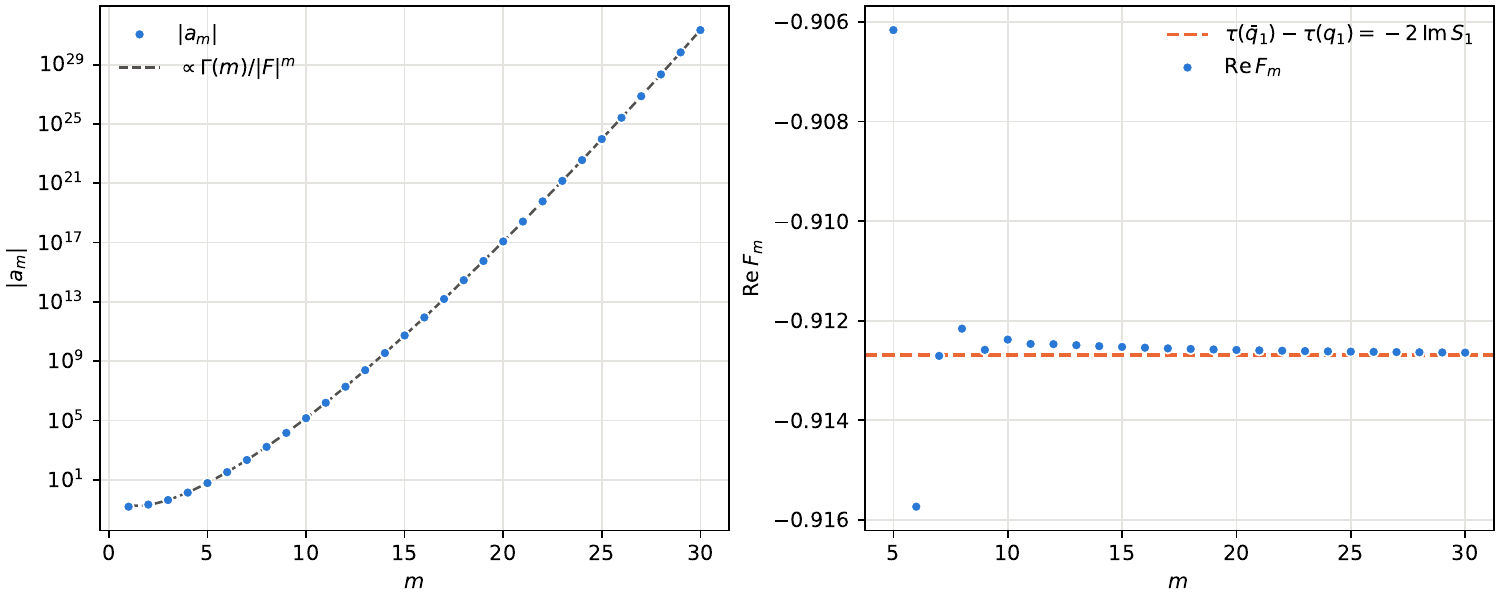}
        \caption{The tail of the series of the complex classical path $q_1$ of the Rosen-Morse barrier ($x_0 = -5$, $x_1 = -1$, $T = 10$). Left: the coefficients $|a_m|$ grow factorially, close to $\Gamma(m)/|F|^m$ (dashed) with $F = -2\,\mathrm{Im}\,S$. Right: the ratio estimate $F_m$ of \cref{eq:Fm} converges to the action difference with the complex-conjugate path $\bar{q}_1$, $F = -0.91269$ (dashed), while its imaginary part fades to $|\mathrm{Im}\, F_{30}| = 1.4 \times 10^{-4}$.}
        \label{fig:ratio}
    \end{figure}

\begin{figure}
        \centering
        \includegraphics[width=0.95\textwidth]{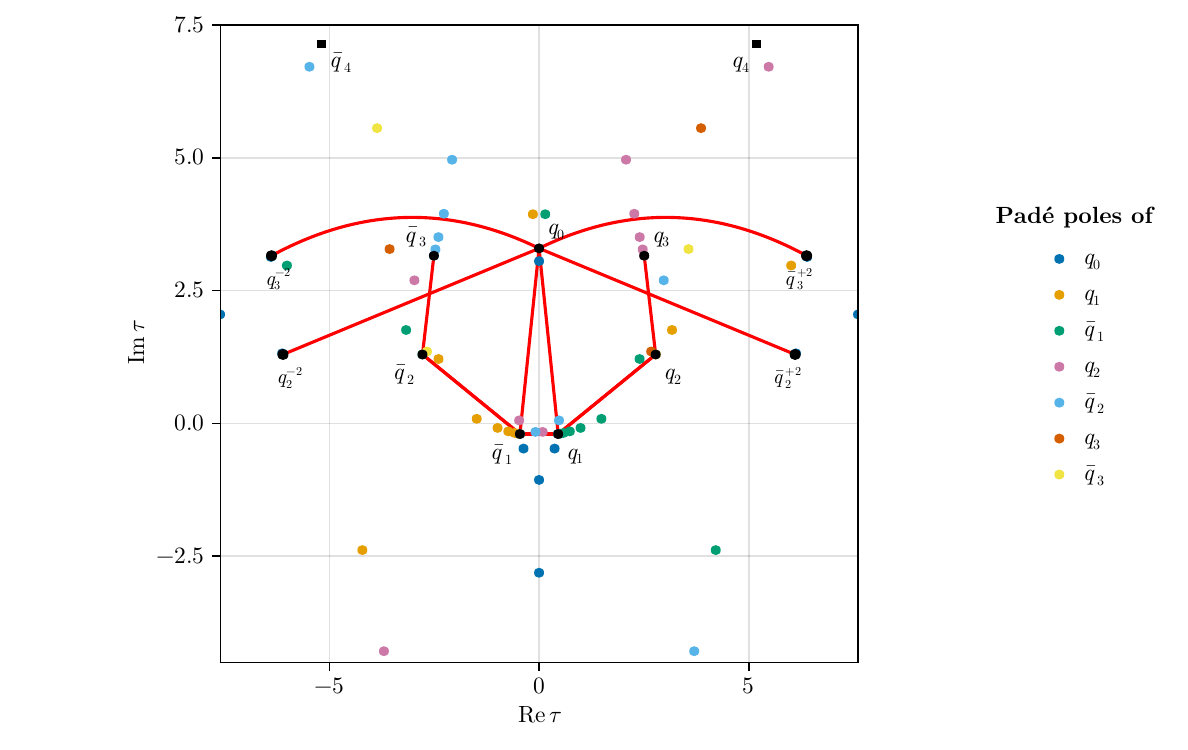}
        \caption{The Borel-Pad\'e poles of the series of the seven computed classical paths of the Rosen-Morse barrier ($x_0 = -5$, $x_1 = -1$, $T = 10$), each shifted by the Borel singularity $\tau_j$ of its own path (black dots). Red lines: the edges with a nonzero Stokes constant. The series of $q_0$ has forty terms, those of $q_1$, $q_2$ and their conjugates thirty, and those of $q_3$ and $\bar{q}_3$ eight. No pole of $q_0$ approaches $q_2$. The outer red lines join $q_0$ to the images $\bar{q}_{2,3}^{+2}$ and $q_{2,3}^{-2}$ (\cref{fig:adjacency}); the edges to $\bar{q}_3^{+2}$ and $q_3^{-2}$ bend to show that they pass by $q_3$ and $\bar{q}_3$. Black squares: the complex paths $q_4$ and $\bar{q}_4$, which end at $x_1 \pm 4\pi\ii$ and whose series we have not computed.}
        \label{fig:poles}
    \end{figure}

    After finding the classical paths, we evaluate the cumulants $\sigma_m^{(j)}$ and the associated series coefficients $a_m^{(j)}$ with the transport recursion relations. The coefficients of the asymptotic series $a_m^{(j)}$ grow factorially with $m$ (see \cref{fig:ratio}). With the complete asymptotic series $a_m^{(j)}$, we formally recover the path integral $G^{(j)}$ with Borel resummation. Evaluate the Borel transform 
    \begin{align}
        \mathcal{B}_j(\tau) = \sum_{m=0}^\infty \frac{T_m^{(j)}\, \tau^m}{m!}
    \end{align}
    and performing a Laplace integral 
    \begin{align}
        G^{(j)}[\bm{x}_1,\bm{x}_0;T] = \left(\frac{1}{2\pi\ii\hbar}\right)^{d/2} e^{\ii S_j/\hbar}\, \int_0^\infty e^{-\tau/\hbar}\, \mathcal{B}_j(\tau)\, \dd \tau / \hbar \,.
    \end{align}
    The Borel transform generally forms a Riemann surface in the Borel plane, the complex plane of the Borel variable $\tau$, distinct from the time $t$ of the classical paths. Each classical path $q_j$ maps onto a branch point singularity at $\tau_j = -\ii S_j$ of the Borel transform in the Borel plane (see \cite{Aniceto:2019,Aniceto:2026,Feldbrugge:2026Wave} for an overview). The real classical paths with real classical actions map onto a branch point singularity on the imaginary axis. The complex classical paths come in complex conjugate pairs, forming pairs of branch point singularities in the Borel plane (see the black points in \cref{fig:poles}). A classical path contributes with the magnitude $|e^{\ii S_j/\hbar}| = e^{-\mathrm{Re}\, \tau_j/\hbar}$. To evaluate the real-time path integral to the accuracy $e^{-\Lambda/\hbar}$, it therefore suffices to study the classical paths with $\mathrm{Re}\, \tau_j = \mathrm{Im}\, S_j \leq \Lambda$.
    
    We number the classical paths by the roots $u_0$ of the boundary condition $\sinh q(T) = \sinh x_1$, in order of increasing $|\mathrm{Im}\, u_0|$: $q_0$ denotes the real path, and $q_1, q_2, \dots$ the complex paths, each with its complex conjugate $\bar{q}_j$; of each pair, $q_j$ has $\mathrm{Re}\, \tau_j > 0$ along the real time axis. For $x_0 = -5$, $x_1 = -1$, and $T = 10$, the path $q_1$ is the relevant complex path, and $q_2$ and $q_3$ end at $x_1 + 2\pi\ii$. Each path also has images. When the time contour passes on the other side of a branch point $t_*$, the action shifts by $\pm\ii\sqrt{2}\,\pi$, the Borel singularity by $\pm\sqrt{2}\,\pi$, and the end point moves between $x_1 + 2\pi\ii n$ and $\ii\pi - x_1 + 2\pi\ii n$; we keep the images that end at $q(T) = x_1$, and denote by $q_j^{k}$ the image at $\tau_j + k\sqrt{2}\pi$. The images $\bar{q}_2^{+2}$ and $\bar{q}_3^{+2}$ of $\bar{q}_2$ and $\bar{q}_3$, each with two passages, lie at $\tau = 6.106 + 1.298\,\ii$ and $\tau = 6.379 + 3.157\,\ii$ (see \cref{fig:adjacency}). Two branch point singularities are adjacent when they can see each other on the Riemann surface of the Borel transform \cite{Dingle:1973,Berry:1991,Howls:1997}. 
    
    This is exactly the information enclosed in the tails of the asymptotic series. In \cref{fig:ratio}, we observe that the series diverges as $\Gamma(m)/|F|^m$ where $F = \tau_l - \tau_j$ is the distance from the singularity $\tau_j$ to its closest adjacent singularity $\tau_l$ in the Borel plane. More concretely, the tail points to a partner. When one adjacent path $q_l$ dominates, the coefficients grow as $a_m^{(j)} \propto \Gamma(m)/F^m$, so that the ratio of consecutive coefficients grows linearly, $a_m^{(j)}/a_{m-1}^{(j)} \approx (m - 1)/F$, and the estimate
    \begin{align}
        F_m = (m - 1)\,\frac{a_{m-1}^{(j)}}{a_m^{(j)}}
        \label{eq:Fm}
    \end{align}
    converges to the distance $F = \tau_l - \tau_j = -\ii(S_l - S_j)$ to the dominant partner. \Cref{fig:ratio} shows $F_m$ for thirty coefficients of the series of $q_1$ from the transport recursion. The estimate settles on the negative real axis, $F_{30} = -0.91264 - 0.00014\,\ii$, with corrections that fade as $1/m$. Which classical path sits there? The complex-conjugate partner $\bar{q}_1$ has the action $S^*$, and $-\ii(S^* - S) = -2\,\mathrm{Im}\, S = -0.91269$. The divergent tail of the complex path, computed from nothing but the path itself and its neighbours, points straight at its mirror image. Going further requires the machinery of resurgence \cite{Dunne:2016,Crew:2026,Aniceto:2026}, and many more terms: to extract the Stokes constant, to subtract the partner's contribution, and to uncover the paths that hide behind it. The transport recursion supplies them. The Stokes constants $K_{jk}\in \mathbb{Z}$ quantify the adjacency ($q_j$ and $q_k$ are adjacent when $K_{jk} \neq 0$), through resurgence relations between the different series expansions
    \begin{align}
        T_m^{(j)} \approx -\frac{K_{jk}}{2\pi\ii} \sum_{l} T_l^{(k)}\, \frac{\Gamma(m - l)}{(\tau_k - \tau_j)^{m - l}}\,.
    \end{align}
    Note that the coefficients $T_m^{(j)}$ are proportional to the coefficients $a_m^{(j)}$.

    With only a finite number of coefficients of each series, we cannot directly implement Borel resummation. To approximate the path integral $G^{(j)}$ and identify the adjacency relations and the associated Stokes constants $K_{jk}$, we employ Borel-Pad\'e methods. We truncate the Borel transform and compute its diagonal Pad\'e approximant $P^{(j)}(\tau)$, a ratio of two polynomials whose poles mimic the singularities. To find the adjacency relations, we evaluate the poles of the Pad\'e approximant and observe that the poles form trails mimicking branch cuts behind adjacent saddle points (see \cref{fig:poles}). To compute the Stokes constant $K_{jk}$ between two branch point singularities $\tau_j$ and $\tau_k$, we perform a conformal map and look for the presence of a pole at a specified location. When the pole is present, its residue yields the Stokes constant (for details see \cite{Aniceto:2026}). To approximate the integral $G^{(j)}$, we evaluate the Laplace transform of the Pad\'e approximant
    \begin{align}
        G^{(j)}[\bm{x}_1,\bm{x}_0;T] \approx \left(\frac{1}{2\pi\ii\hbar}\right)^{d/2} e^{\ii S_j/\hbar}\, \frac{1}{\hbar}\int_0^\infty e^{-\tau/\hbar}\, P^{(j)}(\tau)\, \dd \tau\,.
    \end{align}
    The Borel-Pad\'e approximation is generally exponentially accurate. However, formal error bounds are to the best of the authors' knowledge not currently available. 
    We perform these calculations with the \verb|Julia| package \verb|IntegralResurgence.jl| \cite{Feldbrugge:2026IntegralResurgence}. For the Rosen-Morse path integral, we identify the adjacency graph displayed by the red lines in \cref{fig:poles}. We can also compute the Stokes constants and an exponentially accurate approximation of the path integral using hyperasymptotic methods \cite{Berry:1990,Berry:1991,Howls:1997,OldeDaalhuis:1998,Bennett:2018,Feldbrugge:2026Wave} (also implemented in the \verb|IntegralResurgence.jl| package). In this paper, we restrict, for convenience, the analysis to the Borel-Pad\'e methods.

    \subsection{The relevant classical paths}
\begin{figure}
        \centering
        \includegraphics{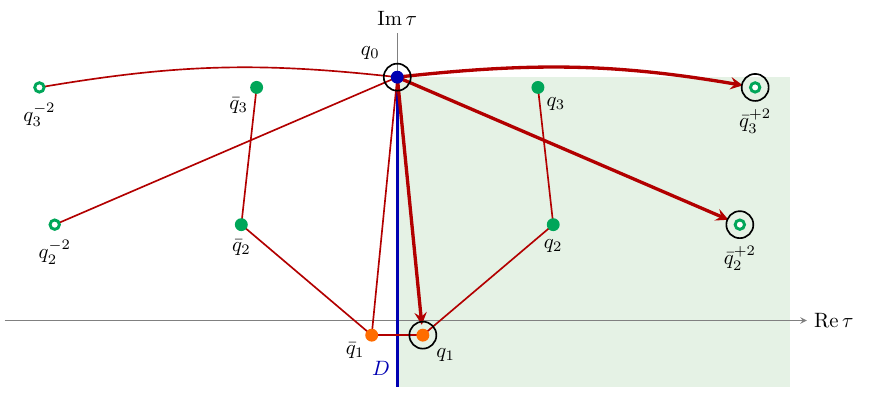}
        \caption{The classical paths of the Rosen-Morse barrier in the Borel plane $\tau = -\ii S$ ($x_0 = -5$, $x_1 = -1$, $T = 10$). Blue: the real path $q_0$; orange: the complex pair $q_1$ and $\bar{q}_1$ born on the fold caustic; green: the complex paths $q_{2,3}$ and their conjugates, which end at $x_1 \pm 2\pi\ii$. The thick blue line marks the image $D$ of the real paths, and the shaded region to its right holds the paths that can contribute. Red lines: adjacency from the Borel-Pad\'e poles, with Stokes constants of unit modulus. Open green circles: the images $\bar{q}_2^{+2}$ and $\bar{q}_3^{+2}$ of $\bar{q}_2$ and $\bar{q}_3$, whose time contours pass above two branch points each and end at $q(T) = x_1$, and their mirror images $q_{2,3}^{-2}$; the poles of the series of $q_0$ reveal their adjacency to $q_0$. Arrows: the edges that the South-East rule fires. Black circles: the contributing paths $q_0$, $q_1$, $\bar{q}_2^{+2}$, and $\bar{q}_3^{+2}$, confirmed by the exact propagator; the images contribute with the weight $e^{-6.1/\hbar}$.}
        \label{fig:adjacency}
    \end{figure}

    We translate the Stokes constants and the geometry of the associated adjacency graph into the intersection numbers using the South-East rule (for more details see \cite{Aniceto:2026}). The real classical paths are always relevant to the path integral. The ascent manifold $\mathcal{K}_j$ of a relevant complex classical path $q_j$ intersects the original integration domain, on which the action is real. Along $\mathcal{K}_j$, the real part $\text{Re}[\ii S/\hbar]$ grows from its value at $q_j$ to zero, so $\text{Re}[\ii S_j] \leq 0$, \textit{i.e.}, $\mathrm{Re}\, \tau_j \geq 0$. Moreover, as the steepest ascent manifold preserves the imaginary part of the exponent, the imaginary part of the exponent $\text{Im}[\ii S_j]$ must match some path in the original integration domain. As the action of Rosen-Morse theory is bounded from below (for real paths), the space of real classical paths maps to a half interval $D$ in the Borel plane, running from the singularity $\tau_0 = -\ii S_0$ of the real classical path with the lowest action to $-\ii \infty$ (see the blue line in \cref{fig:adjacency}). The candidate relevant complex classical paths always reside to the right of the domain $D$ in the Borel plane. This readily excludes many complex classical paths. For a treatment of unbounded actions, we refer to the regulators introduced in \cite{Aniceto:2026}.

    To identify the relevant classical paths, we leverage the adjacency relations. Analytically continuing the path integral $G[\bm{x}_1,\bm{x}_0; T]$ by extending Planck's constant from $(0,\infty)$ to the punctured complex plane $\mathbb{C}\backslash 0$, writing $1/\hbar = e^{\ii \theta}/|\hbar|$, the set of relevant classical paths changes in Stokes phenomena as $\theta$ varies. Varying $\theta$ rotates the adjacency graph around $0$ in the Borel plane. Exact WKB analysis performs the same rotation. As the phase of Planck's constant varies, the Stokes graph of the Schr\"odinger equation deforms, and a Stokes phenomenon occurs when the graph develops a saddle connection, a Stokes curve that joins two turning points. The Delabaere-Dillinger-Pham formula gives the jump \cite{Delabaere:1997}, and the sequence of jumps organises into cluster mutations and the wall-crossing of spectral networks \cite{Iwaki:2014,Gaiotto:2013}. Aoki, Kawai, and Takei traced the Stokes curves back to the steepest descent paths of an integral representation \cite{Aoki:2001}, the link between exact WKB analysis and the thimbles of Picard-Lefschetz theory that the path integral exploits. For $\theta = \pi/2$, the path integral maps to the Euclidean path integral for a particle in a Rosen-Morse well,
    \begin{align}
        \int_{\bm{q}(0)=\bm{x}_0}^{\bm{q}(T)=\bm{x}_1} e^{-S[\bm{q}(t)]/|\hbar|}\mathcal{D}\bm{q}(t)\,.
    \end{align}
    The space of real paths interpolating between $\bm{x}_0$ and $\bm{x}_1$ now maps to the horizontal line $D$ in the Borel plane. The region to the right of $D$ vanishes, \textit{i.e.}, only the real classical paths are relevant. When there are multiple real classical paths, $\theta = \pi/2$ corresponds to a Stokes phenomenon (explained below). We therefore take $\theta = \pi/2 - \epsilon$, with $\epsilon$ small and positive, at which the real classical paths are the relevant ones. When rotating back to $\theta=0$, a complex classical path $q_j$ can become relevant to the path integral 
    \begin{align}
        G^{(j)}_\theta[\bm{x}_1, \bm{x}_0;T] = \int_{\bm{q}(0)=\bm{x}_0}^{\bm{q}(T)=\bm{x}_1} e^{\ii e^{\ii \theta} S[\bm{q}(t)]/|\hbar|}\mathcal{D}\bm{q}(t)\,,
    \end{align}
    in a Stokes phenomenon, when its branch point singularity is horizontally aligned with the branch point of a relevant classical path $q_k$ and the two classical paths are adjacent to each other.
    \begin{figure}
        \centering
        \includegraphics[width=\textwidth]{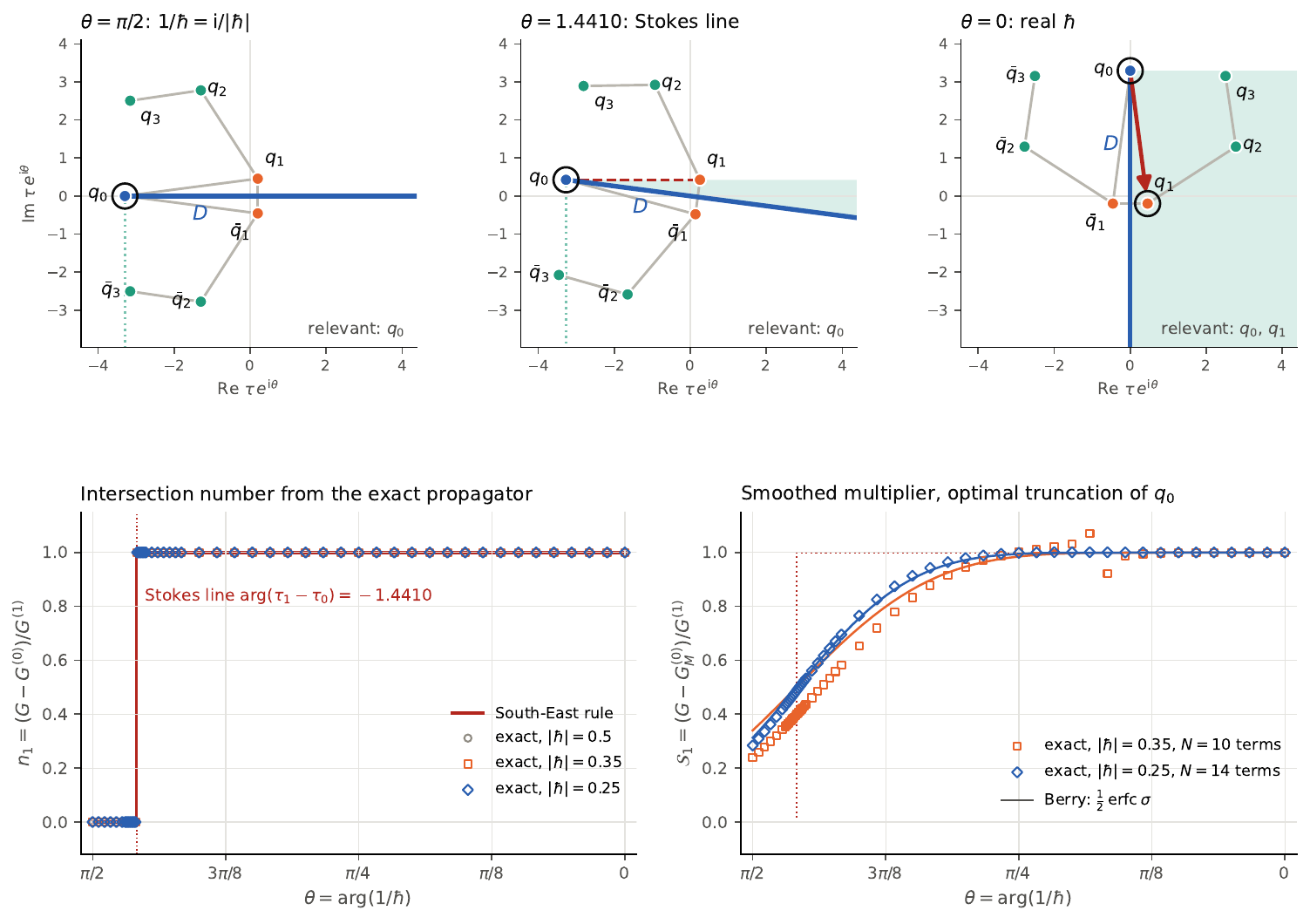}
        \caption{The South-East rule at work for the Rosen-Morse barrier ($x_0 = -5$, $x_1 = -1$, $T = 10$), as $1/\hbar = e^{\ii\theta}/|\hbar|$ rotates from the imaginary to the real axis. Top: the adjacency graph of \cref{fig:adjacency}, rotated by $e^{\ii\theta}$ about the origin of the Borel plane, so that the Laplace ray of $q_0$ always points along the positive real axis. Thick blue: the image $D$ of the real paths; green: the region to the right of $D$ and below the Laplace ray, where a path can contribute; dotted: the lower edge of the South-East quadrant. At $\theta = \pi/2$ the image $D$ lies on the Laplace ray and only $q_0$ contributes. At $\theta = 1.4410$ the edge $q_0 \to q_1$ (red dashed) lies on the Laplace ray: the Stokes line. At $\theta = 0$ the edge has turned into the South-East quadrant and fired (red arrow), and $q_1$ contributes. The panels show the leading paths: the edges from $q_0$ to the images $\bar{q}_{2,3}^{+2}$ of \cref{fig:adjacency} fire at $\theta = 0.316$ and $0.022$, with a weight $e^{-6.1/|\hbar|}$ far below the scale of the plots. Bottom left: the intersection number of $q_1$ from the exact propagator, minus the Borel-Pad\'e sum of $q_0$, divided by the Borel-Pad\'e sum of $q_1$, both along the ray $\arg \tau = -\theta$; red: the South-East rule. Bottom right: the same remainder with $q_0$ truncated after $M \approx |\tau_1 - \tau_0|/|\hbar|$ terms, against Berry's smoothed Stokes multiplier (curves).}
        \label{fig:rotation}
    \end{figure}
    For the Rosen-Morse example, at $\theta = 1.4410$, the complex classical path $q_1$ becomes relevant to the integral (see \cref{fig:rotation}). The branch points of the  classical paths $q_2$ and $q_3$ reside in the green shaded region but remain irrelevant: the real path $q_0$ does not see them, the edge from $q_1$ to $q_2$ points North-East and never fires, and the only edge of $q_3$ leads to the irrelevant path $q_2$. Among the paths of \cref{fig:rotation}, only $q_0$ and $q_1$ are relevant; two images, of exponentially smaller weight, join them below.

    The two lower panels of \cref{fig:rotation} make the Stokes phenomenon visible. Subtract the Borel-Pad\'e sum $G^{(0)}$ of the real path from the exact propagator $G$, and divide the remainder by the Borel-Pad\'e sum $G^{(1)}$ of the complex path, both along the Laplace ray $\arg \tau = -\theta$,
    \begin{align}
        n_1 = \frac{G - G^{(0)}}{G^{(1)}}\,.
    \end{align}
    The ratio measures the intersection number of $q_1$. It vanishes for $\theta > 1.4410$, jumps to one on the Stokes line, and stays one down to real $\hbar$, for every $|\hbar|$ (lower left panel). The sharp step belongs to the Borel sum. Truncate the series of $q_0$ instead at its smallest term, after $M \approx |F|/|\hbar|$ terms with $F = \tau_1 - \tau_0$ giving $G^{(0)}_M$. The remainder of the exact propagator,
    \begin{align}
        \mathcal{S}_1 = \frac{G - G^{(0)}_M}{G^{(1)}}\,,
    \end{align}
    measures a smooth multiplier of $q_1$. Berry predicts its shape \cite{Berry:1989} as an error function across the Stokes line,
    \begin{align}
        \mathcal{S}_1 = \frac{1}{2}\,\mathrm{erfc}\,\sigma\,, \qquad \sigma = \frac{\mathrm{Im}\big(F e^{\ii\theta}/|\hbar|\big)}{\sqrt{2\,\mathrm{Re}\big(F e^{\ii\theta}/|\hbar|\big)}}\,,
    \end{align}
    which narrows as $\sqrt{|\hbar|}$. The lower right panel shows the multiplier, with $\bar{q}_1$ fitted alongside, for $|\hbar| = 0.35$ and $0.25$. It follows Berry's curve, from $1$ at $\theta =0$ to about $0.3$ at $\theta = \pi/2$. The second Stokes line, of $\bar{q}_1$ at $\theta = 1.7007$, lies only $0.130$ away. The optimal truncation smears the jump over a wide range of angles. The Borel sum concentrates it on the Stokes line, where the South-East rule places it.
    
    \begin{figure}
        \centering
        \includegraphics[width=0.7\textwidth]{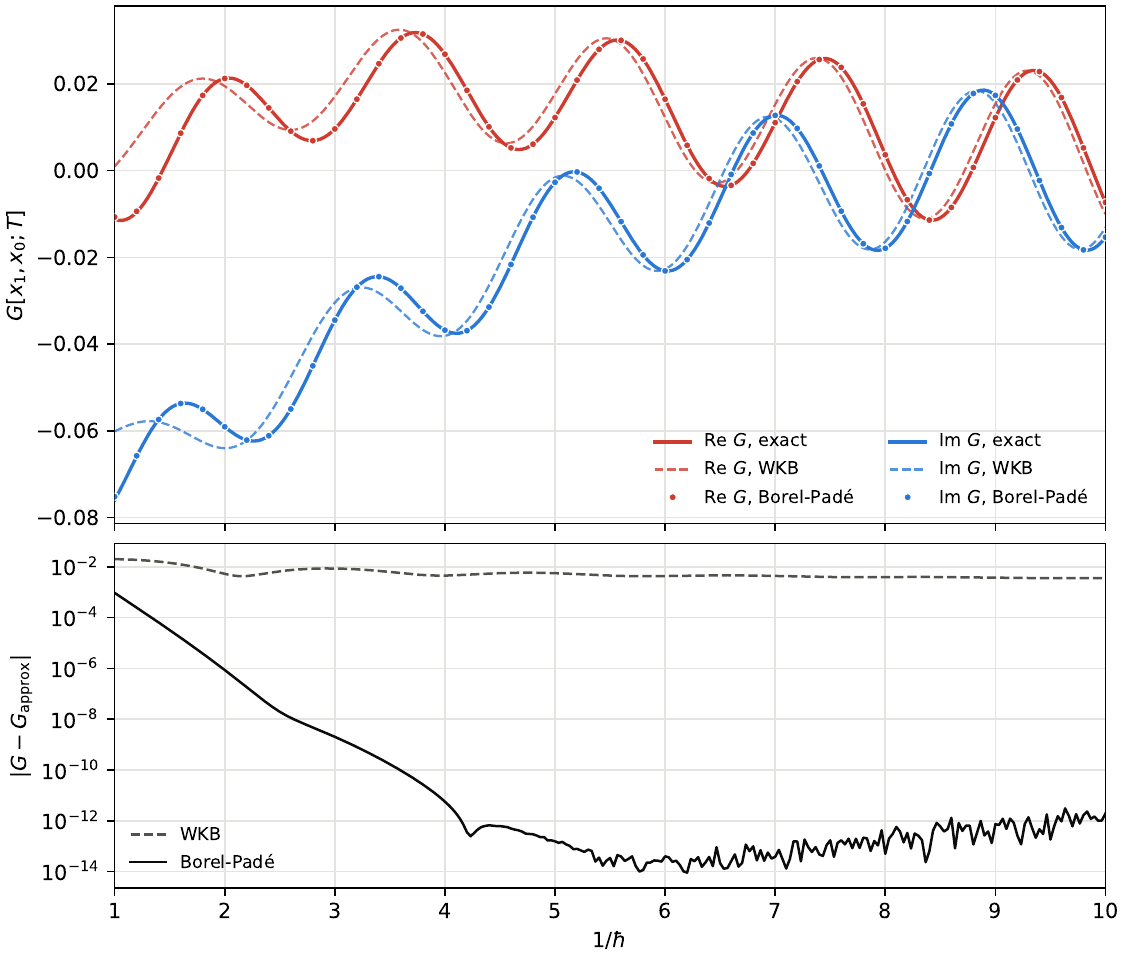}
        \caption{The propagator of the Rosen-Morse barrier ($x_0 = -5$, $x_1 = -1$, $T = 10$) as a function of $1/\hbar$. Top: the real part (red) and the imaginary part (blue) of the exact propagator (solid), of the leading semiclassical approximation of the paths $q_0$ and $q_1$ (dashed), and of the Borel-Pad\'e sums of their series along the ray $\arg \tau = 0$ (dots). Bottom: the error of both approximations. At small $1/\hbar$, the Borel-Pad\'e error follows the images $\bar{q}_{2,3}^{+2}$ of weight $e^{-6.1/\hbar}$ (\cref{fig:adjacency}); at large $1/\hbar$, it levels off at the accuracy of the spectral integral.}
        \label{fig:propagator}
    \end{figure}
   
    Seen from the perspective of the original adjacency graph, we consider all edges running in the South-East direction. Order them from steep to shallow, and update the intersection numbers edge by edge, $n_k \to n_k + n_j K_{jk}$ for the edge $q_j \to q_k$, starting from $n_0 = 1$ for the real classical path. Going back to \cref{fig:adjacency}, in addition to $q_1$, we find the relevant classical paths $\bar{q}_2^{+2}$ and $\bar{q}_3^{+2}$. Their contribution can be detected when evaluating the path integral with the spectral representation. However, their numerical contribution is exponentially suppressed with respect to the contribution of $q_0$ and $q_1$. See \cref{fig:propagator} for a comparison of the exact propagator, the WKB approximation and the Borel-Pad\'e approximation. Remarkably, the classical paths do not only govern the real-time path integral in the semiclassical regime. Using resurgence, we approximate the path integral with exponential accuracy well beyond the semiclassical regime.

    \subsection{Quantum tunnelling and higher-order Stokes phenomena}\label{sec:Tunnelling}
    So far, we considered final points in front of the barrier. We now apply the South-East rule to quantum tunnelling. Keeping the initial point $x_0 = -5$, we increase the time to $T = 15$, and move the final point $x_1$ from $-1$, in front of the barrier, to $4$, behind it. For the longer time, the real classical path lingers on the barrier top, so that its singularity $\tau_0$ rises above the complex classical paths in the Borel plane and the region to the right of $D$ fills with candidates (for $T=10$ and $x_1=-1$ no complex classical path maps to a branch point in the candidate region). For every $x_1$ in the interval, there is a single real classical path. As we move the final point, the singularities associated with the classical paths move in the Borel plane. The relevance of complex classical paths can change in two ways. First, an intersection number changes when an edge enters or leaves the South-East sector, or when two edges change their order of firing. This is an ordinary Stokes phenomenon, in which the final point rather than the phase of $\hbar$ moves the edge across a Stokes line. Second, the adjacency graph can change in a higher-order Stokes phenomenon \cite{Howls:2004,Aniceto:2026,Crew:2026}, when a singularity $\tau_k$ crosses the line segment between $\tau_i$ and $\tau_j$, while $q_i$ is adjacent to $q_k$ and $q_k$ to $q_j$. In the exact WKB analysis of higher-order differential equations, the same phenomenon appears as the new Stokes curves of Berk, Nevins, and Roberts, which emanate from virtual turning points rather than ordinary ones \cite{Berk:1982,Honda:2015,Howls:2004}. 
    
    \begin{figure}
        \centering
        \includegraphics[width=0.9\textwidth]{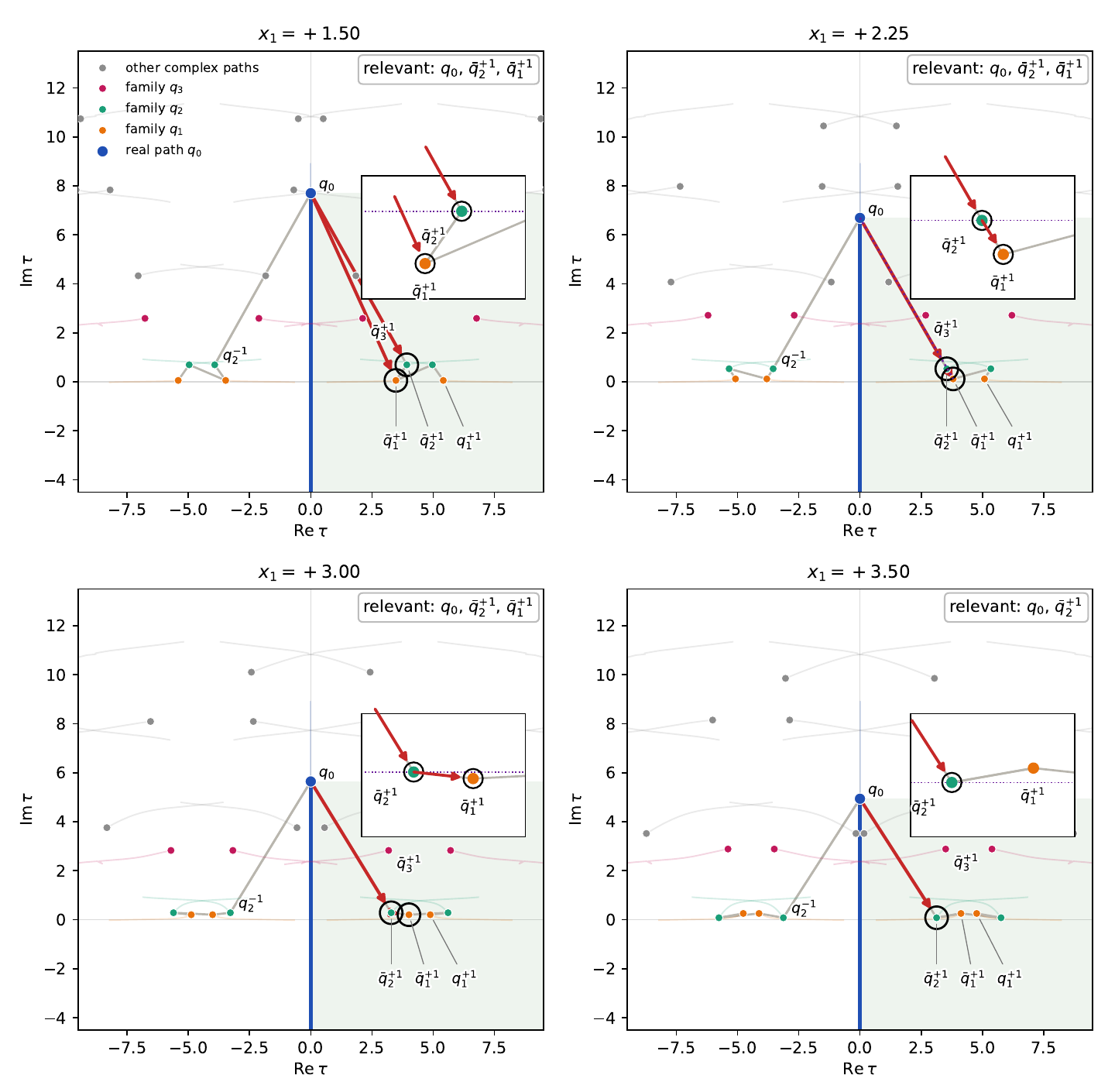}
        \caption{The Borel plane of the Rosen-Morse barrier ($x_0 = -5$, $T = 15$) at four final points: every classical path and image that ends at $x_1$, within a distance $9$ of the real path $q_0$ (large blue dot), coloured by family, with faded trails. The shaded region to the right of $D$ holds the paths that can contribute; circles mark the relevant paths. Grey lines: the adjacency graph, computed at $x_1 = 4$ from the series of the paths and images near $q_0$ and their mirrors, and carried along the tracks; it changes only in higher-order Stokes phenomena. Red arrows: the edges that fire in the South-East rule. The insets magnify the neighbourhood of $\bar{q}_2^{+1}$ and $\bar{q}_1^{+1}$, with the line $\arg = 0$ through $\bar{q}_2^{+1}$. At $x_1 = 1.5$, the path $\bar{q}_1^{+1}$ draws its weight from $q_0$ directly. At $x_1 = 2.25$, the singularities of $q_0$, $\bar{q}_2^{+1}$, and $\bar{q}_1^{+1}$ line up, a higher-order Stokes phenomenon, and the weight passes to the chain $q_0 \to \bar{q}_2^{+1} \to \bar{q}_1^{+1}$. At $x_1 = 3.0$, the edge $\bar{q}_2^{+1} \to \bar{q}_1^{+1}$ still points South-East; at $x_1 = 3.5$, it points North-East, and $\bar{q}_1^{+1}$ drops out.}
        \label{fig:sweep}
    \end{figure}

    To track these changes, we follow every root $u_0(x_1)$ of the boundary condition $\sinh q(T) = \sinh x_1$ as a continuous function of $x_1$ (see \cref{fig:sweep}). The South-East rule explains the transitions in two steps. At $x_1 = 1.5$, the edge $q_0 \to \bar{q}_1^{+1}$ points South-East, and $\bar{q}_1^{+1}$ is relevant, while the edge $\bar{q}_2^{+1} \to \bar{q}_1^{+1}$ lies outside the South-East sector. As $x_1$ increases, the edge $\bar{q}_2^{+1} \to \bar{q}_1^{+1}$ rotates towards the real axis, until, at $x_1 = 2.25$, the singularities $\tau_0$, $\tau_{\bar{q}_2^{+1}}$, and $\tau_{\bar{q}_1^{+1}}$ are collinear. In this higher-order Stokes phenomenon, the Stokes constant of the edge $q_0 \to \bar{q}_1^{+1}$ drops from one to zero, and the chain $q_0 \to \bar{q}_2^{+1} \to \bar{q}_1^{+1}$ takes over its role, so that the intersection number of $\bar{q}_1^{+1}$ remains one. At $x_1 \approx 3.16$, where $\mathrm{Im}\, \tau_{\bar{q}_1^{+1}} = \mathrm{Im}\, \tau_{\bar{q}_2^{+1}}$, the edge $\bar{q}_2^{+1} \to \bar{q}_1^{+1}$ leaves the South-East sector through the real direction. Here, the Borel sum of $\bar{q}_2^{+1}$ absorbs the contribution of $\bar{q}_1^{+1}$, and $\bar{q}_1^{+1}$ becomes irrelevant. Fits of the exact propagator confirm both steps: $n_{\bar{q}_1^{+1}} \approx 1$ up to $x_1 = 3.1$ and $n_{\bar{q}_1^{+1}} \approx 0$ from $x_1 = 3.25$ on, while $n_{\bar{q}_2^{+1}} = 1$ throughout. In summary, the adjacency graph changes once, in the higher-order Stokes phenomenon, and the intersection numbers change once more, in an ordinary Stokes phenomenon.

    \begin{figure}
        \centering
        \includegraphics[width=0.7\textwidth]{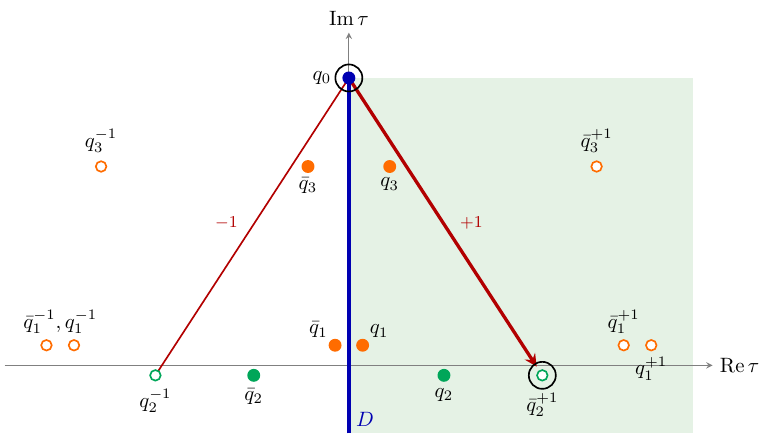}
        \caption{The classical paths of the Rosen-Morse barrier behind the barrier at the long time $T = 15$, in the Borel plane $\tau = -\ii S$ ($x_0 = -5$, $x_1 = 4$). Blue: the real path $q_0$, which lingers on the barrier top and lies above the complex paths. Filled: the complex paths along the straight time contour, which end at an image of $x_1$: the pair $q_1$, $\bar{q}_1$ and the pair $q_3$, $\bar{q}_3$ with nearly imaginary initial velocity (orange), and the below-barrier pair $q_2$, $\bar{q}_2$ (green). Open: their images around one branch point of $q(t)$, which reach $x_1$: $\bar{q}_2^{+1}$, $q_2^{-1}$ from $\bar{q}_2$, $q_2$; $\bar{q}_3^{+1}$, $q_3^{-1}$ from $\bar{q}_3$, $q_3$; and the pair $q_1^{+1}$, $\bar{q}_1^{+1}$ from $q_1$, $\bar{q}_1$, with their mirrors. The shaded region to the right of $D$ holds the paths that can contribute. Red lines: the edges of $q_0$ with the Stokes constants from its large-order growth; the arrow marks the edge $q_0 \to \bar{q}_2^{+1}$ that fires. Circles: the contributing paths $q_0$ and $\bar{q}_2^{+1}$, confirmed by the exact propagator.}
        \label{fig:tunnel15}
    \end{figure}

    At the final point $x_1 = 4$, we find two relevant classical paths (see \cref{fig:tunnel15}). The real classical path $q_0$ lingers at the top of the barrier before reaching $x_1$. The edge $q_0 \to \bar{q}_2^{+1}$, with the Stokes constant $K = +0.995$ from the large-order growth of the series of $q_0$, points South-East and fires, and the exact propagator confirms $n_{\bar{q}_2^{+1}} = 1$ to $4 \times 10^{-4}$. The complex classical path $\bar{q}_2^{+1}$ captures quantum tunnelling. This complex classical path crosses the barrier with an energy far below its top and reaches $x_1$ around a singularity of the potential close to the real time axis. The images $q_1^{+1}$ and $\bar{q}_1^{+1}$ of $q_1$ and $\bar{q}_1$, which capture the tunnelling rate of the energy propagator \cite{Feldbrugge:2025RosenMorse}, do not contribute to the propagator at this final point.

    \begin{figure}
        \centering
        \includegraphics[width=0.7\textwidth]{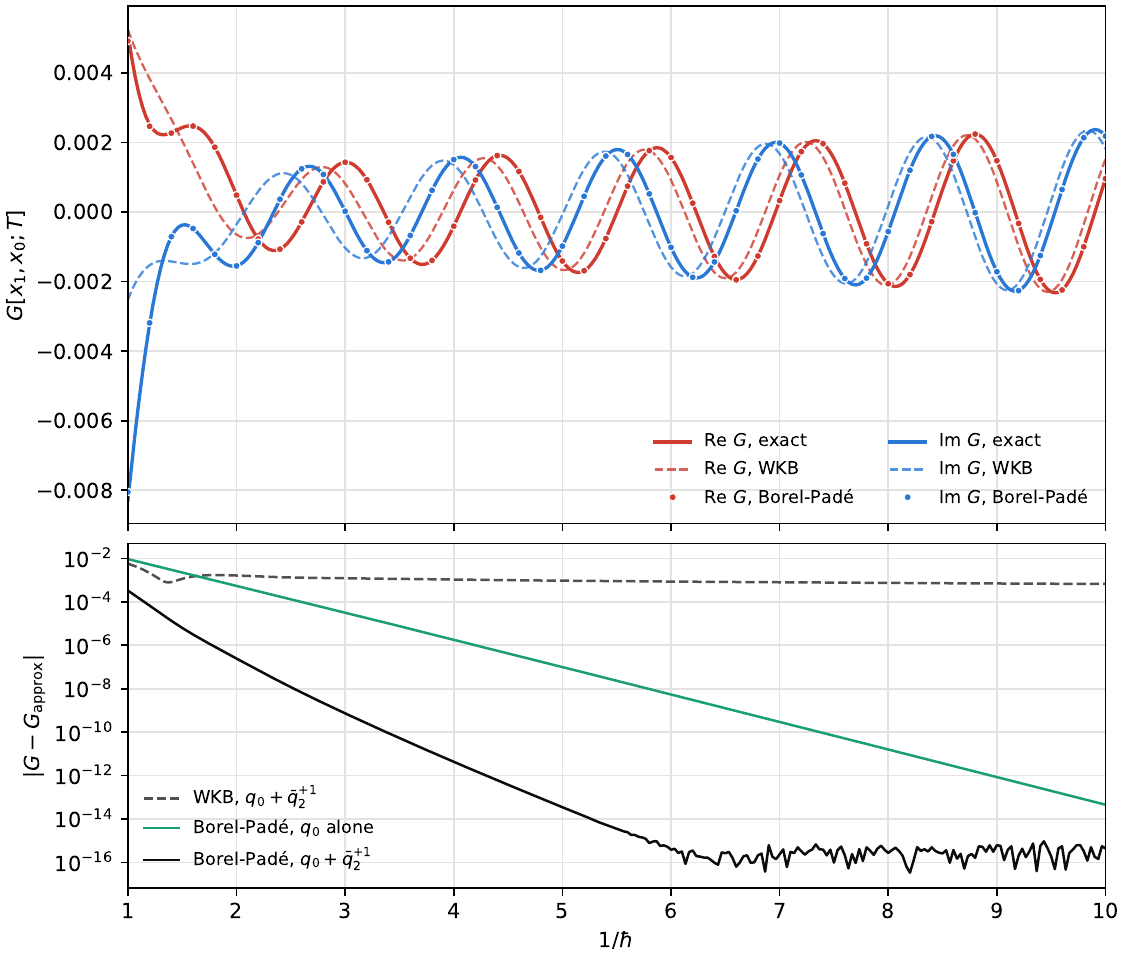}
        \caption{The propagator of the Rosen-Morse barrier behind the barrier ($x_0 = -5$, $x_1 = 4$, $T = 15$) as a function of $1/\hbar$. Top: the real part (red) and the imaginary part (blue) of the exact propagator (solid), of the leading semiclassical approximation of the paths $q_0$ and $\bar{q}_2^{+1}$ (dashed), and of the Borel-Pad\'e sums of their series along the ray $\arg \tau = 0$ (dots). Bottom: the error of the leading approximation, of the Borel-Pad\'e sum of the real path $q_0$ alone, which falls as $e^{-2.98/\hbar}$, the weight of the tunnelling path $\bar{q}_2^{+1}$, and of the Borel-Pad\'e sums of $q_0$ and $\bar{q}_2^{+1}$, which levels off at the accuracy of the spectral integral.}
        \label{fig:propagatortunnel}
    \end{figure}

    \Cref{fig:propagatortunnel} compares the exact propagator with the semiclassical approximations for $1 \leq 1/\hbar \leq 10$. The leading semiclassical approximation of $q_0$ and $\bar{q}_2^{+1}$ deviates from the exact propagator by a third of its magnitude, and the error decreases only slowly with $\hbar$. As the real classical path lingers on the barrier top, its quantum corrections $a_m^{(0)}\hbar^m$ are large compared with the leading term $a_0^{(0)} = 1$. The Borel-Pad\'e approximation of $q_0$ alone is far more accurate, but its error decreases as $e^{-2.98/\hbar}$, the weight of $\bar{q}_2^{+1}$, leaving an error of $5 \times 10^{-14}$ at $1/\hbar = 10$. Including the Borel-Pad\'e approximation of $\bar{q}_2^{+1}$ reduces the error to $4 \times 10^{-16}$ by $1/\hbar = 6$, the accuracy of the spectral integral. With the resummed series of both classical paths, the tunnelling path is as visible in the propagator as the real classical path.

    \section{The relativistic path integral}\label{sec:Relativistic}
    So far, we considered the relevance of classical paths in the non-relativistic path integral. We here briefly comment on how these methods extend to relativistic theories governed by a constrained equation. An upcoming paper applies resurgence theory and the South-East rule to relativistic theories in detail. 

    In relativistic quantum systems, reparametrisation invariance replaces the Schr\"odinger equation by a constrained equation 
    \begin{align}
        \hat{H} \psi(\bm{x}) = 0\,,
    \end{align}
    as in Klein-Gordon theory (with $\bm{x}$ a space-time event) \cite{Feynman:1950,Feynman:1951,Schwinger:1951} or for the Wheeler-DeWitt equation of quantum geometrodynamics (with $\bm{x}$ a three-geometry in superspace) \cite{Wheeler:1957,Wheeler:1968,DeWitt:1967,DeWitt:1967b,DeWitt:1967c}. In both cases, the constraint takes the form
    \begin{align}
        \hat{H} = -\frac{\hbar^2}{2}\,g^{AB}\nabla_A\nabla_B + U(\bm{x})\,, 
    \end{align}
    on a configuration space with the coordinates $x^A$, $A = 1, \dots, D$, the metric $g_{AB}$, its covariant derivative $\nabla_A$, and the Laplace-Beltrami operator $g^{AB}\nabla_A\nabla_B$. For the Klein-Gordon equation, $\bm{x}$ collects the spacetime coordinates, $g_{AB}$ forms the spacetime metric, and the potential $U = m^2/2$, possibly with an external potential. For the Wheeler-DeWitt equation in minisuperspace, $\bm{x}$ collects the three-geometry and the homogeneous matter fields, $g_{AB}$ forms the DeWitt supermetric, and $U$ collects the curvature of the spatial slices, the cosmological constant, and the potential of the matter fields. In both cases, the metric carries a negative direction, and no external clock appears. The operator $\hat{H}$ thus takes the form of a non-relativistic Hamiltonian with unit mass on a curved configuration space with an indefinite metric. 

    \bigskip

    The transport recursion method for a non-relativistic particle relies on the time parameter, which is absent for relativistic propagators. The world-line quantisation method restores the clock in the form of the lapse
    \begin{align}
        G[\bm{x}_1, \bm{x}_0] = \int_C G[\bm{x}_1, \bm{x}_0; N]\, \mathrm{d}N\,, \label{eq:lapse}
    \end{align}
    where $G[\bm{x}_1, \bm{x}_0; N]$ satisfies the Schr\"odinger-like equation
    \begin{align}
        \ii \hbar \frac{\partial G[\bm{x}_1, \bm{x}_0; N]}{\partial N} = \hat{H} G[\bm{x}_1, \bm{x}_0; N]\,. \label{eq:lapsekernel}
    \end{align}
    Whereas in quantum field theory, $N$ goes by the name of Schwinger's proper time \cite{Feynman:1950,Schwinger:1951}, in quantum gravity it is known as the lapse \cite{Teitelboim:1980,Teitelboim:1982,Halliwell:1988}. The contour $C = (0, \infty)$ implements causality and yields the Feynman propagator, a Green's function of the constraint. The contour $C = \mathbb{R}$ implements reparametrisation invariance and yields a solution of the constrained equation \cite{Teitelboim:1983}.

    At a fixed lapse $N$, we compute the semiclassical expansion of $G[\bm{x}_1,\bm{x}_0;N]$ with the transport recursion method. With the endpoints held fixed, each thimble of the kernel takes the form
    \begin{align}
        G^{(j)}[\bm{x}_1, \bm{x}_0; N] \sim e^{\frac{\ii S_j(N)}{\hbar} + \alpha^{(j)}(N) + \sum_{m=1}^\infty \sigma_m^{(j)}(N)\, \hbar^m}\,.
    \end{align}
    We obtain the full semiclassical expansion with an additional saddle-point expansion around the stationary points $N_s$ of the lapse integral. At fixed endpoints, the stationary points $N_s$, where $\partial_N S_j = 0$, select the classical paths that solve the classical constraint. Below, all derivatives $\partial_N$ act at fixed endpoints and are evaluated at $N_s$. The Gaussian integral over $N - N_s$ yields the expansion 
    \begin{align}
        G^{(j)}[\bm{x}_1, \bm{x}_0] \sim 
        \left(\frac{1}{2\pi \ii \hbar}\right)^{(D-1)/2} \frac{1}{\left|\det g_{AB}(\bm{x}_0) \det g_{AB}(\bm{x}_1)\right|^{1/4}} \sqrt{\frac{ \det (-\partial^2 S_j / \partial \bm{x}_0 \partial \bm{x}_1)}{\partial_N^2 S_j}}
        e^{\ii S_j(N_s)/\hbar} e^{\sum_{m=1}^\infty \tilde{\sigma}_m^{(j)} \hbar^m}\,,
    \end{align}
    with the determinant evaluated at $N_s$. The factors $|\det g_{AB}|^{-1/4}$ turn the density $\det(-\partial^2 S_j/\partial\bm{x}_0\partial\bm{x}_1)$ into a scalar, as befits a propagator normalised as a scalar at both endpoints, $\psi(\bm{x}_1) = \int G[\bm{x}_1, \bm{x}_0]\, \psi(\bm{x}_0) \sqrt{|\det g_{AB}(\bm{x}_0)|}\, \dd \bm{x}_0$, and drop out for a constant metric \cite{DeWitt:1957}. The cumulants $\tilde{\sigma}_m^{(j)}$ of the propagator differ from the cumulants $\sigma_m^{(j)}(N_s)$ of the kernel, as the integral over the lapse adds corrections. The Feynman rules of \cref{sec:Diagrams}, now in a single dimension with the line $\ii\hbar/\partial_N^2 S_j$, yield the first modified cumulant,
    \begin{align}
        \tilde{\sigma}_1^{(j)} = \sigma_1^{(j)}(N_s) + \ii\left[\frac{5\,(\partial_N^3 S_j)^2}{24\,(\partial_N^2 S_j)^3} - \frac{\partial_N^4 S_j}{8\,(\partial_N^2 S_j)^2} - \frac{\partial_N^3 S_j\,\partial_N\alpha^{(j)}}{2\,(\partial_N^2 S_j)^2} + \frac{\partial_N^2\alpha^{(j)} + (\partial_N\alpha^{(j)})^2}{2\,\partial_N^2 S_j}\right]\,.
    \end{align}

    The constraint offers a second road to the cumulants $\tilde{\sigma}_m^{(j)}$. Away from the initial point, the propagator solves the equation $\hat{H} G^{(j)}[\bm{x}_1, \bm{x}_0] = 0$. Substituting the ansatz $G^{(j)} = A_j e^{\ii S_j/\hbar + \sum_{m=1}^\infty \tilde{\sigma}_m^{(j)} \hbar^m}$, with the action  $S_j$ at the stationary point, and sorting by powers of $\hbar$, we obtain the Hamilton-Jacobi constraint
    \begin{align}
        \frac{1}{2} g^{AB} \partial_A S_j\, \partial_B S_j + U = 0\,.
    \end{align}
    At the next order, the logarithm of the amplitude, $\tilde{\alpha}^{(j)} = \ln A_j$, obeys the transport equation
    \begin{align}
        \frac{\dd \tilde{\alpha}^{(j)}}{\dd \lambda} = -\frac{1}{2}\, g^{AB}\nabla_A\nabla_B S_j\,,
    \end{align}
    and the remaining orders yield the constrained transport recursion
    \begin{align}
        \frac{\dd \tilde{\sigma}_1^{(j)}}{\dd \lambda} &= \frac{\ii}{2}\, g^{AB}\left[\nabla_A\nabla_B \tilde{\alpha}^{(j)} + \partial_A\tilde{\alpha}^{(j)}\, \partial_B\tilde{\alpha}^{(j)}\right],\nonumber\\
        \frac{\dd \tilde{\sigma}_{m+1}^{(j)}}{\dd \lambda} &= \frac{\ii}{2}\, g^{AB}\left[\nabla_A\nabla_B \tilde{\sigma}_m^{(j)} + 2\,\partial_A\tilde{\alpha}^{(j)}\, \partial_B\tilde{\sigma}_m^{(j)} + \sum_{k=1}^{m-1} \partial_A\tilde{\sigma}_k^{(j)}\, \partial_B\tilde{\sigma}_{m-k}^{(j)}\right],
        \label{eq:constrainedtransport}
    \end{align}
    with the derivative $\frac{\dd}{\dd \lambda} = g^{AB}\, \partial_A S_j\, \partial_B$ along the classical path, whose velocity reads $\frac{\dd x^A}{\dd\lambda} = g^{AB}\partial_B S_j$. The constrained recursion mirrors the transport recursion \eqref{eq:transport} with unit mass: the constraint carries no clock, so the partial time derivative drops out of the co-moving derivative, and the curved metric $g^{AB}$ replaces the flat one. These constrained transport relations yield the same cumulants as the integral over the lapse. However, they lack the initial conditions which completed the system for the non-relativistic particle. The world-line method supplies them, as the kernel starts from a delta function at $N = 0$, which fixes $\sigma_m^{(j)}(0) = 0$, and the integral over the lapse imposes the constraint only at the end.

    For a single degree of freedom, as in a minisuperspace model with only the scale factor, the constraint reduces to a time-independent Schr\"odinger equation at zero energy, the home ground of exact WKB analysis. Exact WKB analysis already describes Schwinger pair production in time-dependent electric fields and particle production in an expanding universe as Stokes phenomena of the mode equations \cite{Dumlu:2010,Taya:2021,Enomoto:2021}. The lapse integral offers a path integral road to the same Stokes phenomena, one that extends to several degrees of freedom.

    \FloatBarrier
    \section{Conclusion}\label{sec:Conclusion}
    To our knowledge, this paper uses resurgence for the first time to identify the relevant complex classical paths of the non-relativistic real-time Feynman path integral. Every classical path, real or complex, carries a divergent series in Planck's constant, and we compute these series to high order, with a Feynman-diagram expansion and with a transport recursion for the cumulants. Their divergent tails encode the adjacency of the classical paths: the late coefficients of one series point to the singularities of its neighbours in the Borel plane. Remarkably, Dingle's observation, developed for finite-dimensional integrals and ordinary differential equations, appears to carry over to the infinite-dimensional path integral, with the classical paths in the role of the saddle points. Continued into the complex plane, classical mechanics thus encodes quantum physics: the relevant classical paths, together with their neighbouring paths, fix the path integral to exponential accuracy, deep in the quantum regime as well as in the semiclassical limit.

    The South-East rule turns adjacency into relevance. Starting from the Euclidean path integral, in which only the real classical paths contribute, we rotate the phase of Planck's constant back to the real axis, and switch on a complex classical path whenever an edge from a relevant path sweeps across its Laplace ray. With Borel-Pad\'e methods, and in principle with hyperasymptotics, this procedure identifies the relevant complex classical paths, the instantons, without computing a single steepest ascent manifold in the infinite-dimensional space of paths. For the Rosen-Morse barrier, the spectral representation of the exact propagator confirms every verdict. It confirms the relevant complex path in front of the barrier, the images that contribute at exponentially smaller weight, the tunnelling path behind the barrier, and the switching of the relevant paths as the final point moves through ordinary and higher-order Stokes phenomena. The classical paths do more than govern the semiclassical limit. The leading semiclassical approximation captures the propagator only as Planck's constant tends to zero, yet the Borel sums of the series of the relevant classical paths approximate the real-time path integral at any value of Planck's constant. The error of a finite set of relevant paths follows the exponential weight of the first path left out, and including that path pushes the error down to the next. Behind the Rosen-Morse barrier, the leading semiclassical approximation misses the propagator by a third of its magnitude, while the Borel-Pad\'e sums of the real and the tunnelling path reproduce it to the accuracy of the spectral integral from $1/\hbar = 6$ on. In front of the barrier, the remaining error of the Borel-Pad\'e sums follows the images of exponentially smaller weight.
    
    The present methods fail at caustics. As the final point approaches a caustic, two classical paths coalesce: the determinant of the Jacobi matrix $Y(T)$ vanishes, the amplitudes of both paths diverge, and their Borel singularities approach each other, so that the late coefficients of each series blow up as $\Gamma(m)/F^m$ with a shrinking distance $F$. Near a caustic, the separate series of the coalescing paths no longer describe the propagator. Uniform asymptotic expansions may resolve this failure. They treat the coalescing paths as one, replacing their pair of Gaussian approximations with an Airy function, or with a higher catastrophe integral at a higher-order caustic \cite{Chester:1957,Berry:1980}. The hyperasymptotics of degenerate saddle points and the uniform WKB method show how to combine such expansions with resurgence \cite{Bennett:2018,Dunne:2014}.

    These results provide evidence, not a formal proof. The real-time path integral as a genuine integral over paths defined in terms of a $\sigma$-measure lacks a generally accepted rigorous definition. The construction of Feldbrugge and Turok, as the continuum definition in terms of absolutely convergent integrals over Lefschetz thimbles, suggests rather than establishes that such a definition exists \cite{Feldbrugge:2023}. A proof that the South-East rule selects the relevant classical paths of the path integral, and that the Borel sums of their series reproduce it, remains an open problem. The agreement with the exact Rosen-Morse propagator, to exponential accuracy and through ordinary and higher-order Stokes phenomena, supports both claims. The higher-order Stokes phenomena also suggest an economy. In a future paper, we will explore whether it suffices to compute the adjacency graph in one well-chosen configuration and propagate the adjacency relations to other configurations using higher-order Stokes phenomena.

    Excitingly, this work suggests that we could define the path integral in terms of the transseries without referring to a proper integral over paths. The transport recursion relations provide the transseries up to the intersection numbers. The South-East rule fixes them through the resurgence adjacency relations. In such a construction, the path integral (and Green's function) is defined by the South-East rule rather than satisfying it. Ideally, we would derive the South-East rule from the definition of the Green's function. While this is an exciting algebraic proposal, it fails to explain the link to the interference of possible histories which inspired Feynman and Wheeler to construct the path integral and provides us valuable intuitions into quantum physics.

    The adjacency graph plays the role of the Stokes graph of exact WKB analysis \cite{Voros:1983,Delabaere:1997,Kawai:2005}. Both record which contributions see each other in the Borel plane, both carry Stokes constants, of unit modulus in the simplest cases, and both change in higher-order Stokes phenomena \cite{Berk:1982,Howls:2004,Honda:2015}. Exact WKB analysis reads the graph from the turning points of an ordinary differential equation. The present method reads it from the divergent series of the classical paths, and therefore applies to the infinite-dimensional path integral.

    The same question arises wherever complex classical solutions enter. In quantum gravity, Kontsevich and Segal, and Witten, have argued for criteria on the allowed complex metrics \cite{Witten:2010,Witten:2021,Kontsevich:2021}, and the recent computations of the entropy of Hawking radiation and the Page curve rest on the contribution of complex saddle points, the replica wormholes \cite{Penington:2020,Almheiri:2019,Almheiri:2020,Penington:2022}. The method presented here offers a way to decide which of these saddle points contribute. As a first step, \cref{sec:Relativistic} comments on the extension of the semiclassical expansion to the Hamiltonian constraints of relativistic quantum mechanics and quantum cosmology. Quantum field theory and quantum gravity remain the long-term goals. 

    \acknowledgments
    The author thanks Ines Aniceto, Chris Howls, Ue-Li Pen and Samuel Crew for many illuminating discussions and collaborations over the last years that lead up to this paper. The author acknowledges the use of Claude Opus 5.5 for assistance with presentation and drafting, figure preparation, and numerical prototyping and cross-checks. All mathematical claims and results are the responsibility of the author. The author is supported by a Postdoctoral Fellowship at National Taiwan University funded by the National Science and Technology Council (NSTC) 114-2923-M-002-011- MY5 Max Planck-IAS-NTU Center for Particle Physics, Cosmology and Geometry.

    \bibliography{Library}

    \appendix
    \section{The spectral representation of Rosen-Morse theory}\label{ap:RosenMorse}
    The Rosen-Morse barrier \cite{Rosen:1932}, also known as the P\"oschl-Teller potential \cite{Poschl:1933}, belongs to the few potentials whose energy eigenstates are known in closed form \cite{Eckart:1930,Rosen:1932,Poschl:1933}. Summed over these eigenstates, the spectral representation yields an exact expression for the propagator. Crandall evaluated it for the reflectionless wells of this family \cite{Crandall:1983}, and Kleinert and Mustapic summed it for the fixed-energy amplitudes of the P\"oschl-Teller and Rosen-Morse potentials \cite{Kleinert:1992}. We follow the evaluation of the real-time propagator of the barrier in Ref.~\cite{Feldbrugge:2025RosenMorse}, and compare it with the semiclassical expansion. For the potential $V(x) = 1/\cosh^2 x$ with unit mass, the time-independent Schr\"odinger equation with the energy $E = \hbar^2 k^2/2$ is solved by the conical functions
    \begin{align}
        \phi_k^{\pm}(x) = \sqrt{\frac{k \sinh(\pi k)}{\cosh(2\pi k) + \cosh(2\pi\nu)}}\, P^{\ii k}_{-\frac{1}{2} + \ii\nu}(\pm\tanh x)\,, \qquad \nu = \frac{\sqrt{8 - \hbar^2}}{2\hbar}\,,
    \end{align}
    with $k > 0$ and the associated Legendre functions $P^{\mu}_{\lambda}$. For $\hbar^2 < 8$, the parameter $\nu$ is real. The states $\phi_k^{+}$ and $\phi_k^{-}$ describe a particle incident from the left and from the right, and they are normalised as $\int \phi^{j_1}_{k_1}(x)\, \phi^{j_2}_{k_2}(x)^*\, \dd x = \delta_{j_1 j_2}\, \delta(k_1 - k_2)$ \cite{Feldbrugge:2023b}. The spectral representation of the propagator reads
    \begin{align}
        G[x_1, x_0; T] = \Theta(T) \int_0^\infty \sum_{j = \pm} \phi^{j}_k(x_1)\, \phi^{j}_k(x_0)^*\, e^{-\ii\hbar k^2 T/2}\, \dd k\,.
        \label{eq:spectral}
    \end{align}
    Expressing the Legendre functions in terms of hypergeometric functions, the products of the eigenstates take the form
    \begin{align}
        \phi^{j}_k(x_1)\, \phi^{j}_k(x_0)^* = \;&\frac{1}{\pi}\, \frac{\sinh^2(\pi k)}{\cosh(2\pi k) + \cosh(2\pi\nu)}\, e^{\ii j k (x_1 - x_0)} \nonumber\\
        &\times {}_2F_1\!\left(\tfrac{1}{2} - \ii\nu, \tfrac{1}{2} + \ii\nu; 1 - \ii k; \tfrac{1 - j \tanh x_1}{2}\right)
        {}_2F_1\!\left(\tfrac{1}{2} - \ii\nu, \tfrac{1}{2} + \ii\nu; 1 + \ii k; \tfrac{1 - j \tanh x_0}{2}\right),
    \end{align}
    with $j = \pm 1$ \cite{Feldbrugge:2025RosenMorse}.

    Along the real $k$ axis, the integral \eqref{eq:spectral} oscillates and converges only conditionally. We therefore deform the integration domain into the complex $k$-plane \cite{Feldbrugge:2023}, where the factor $e^{-\ii\hbar k^2 T/2}$ decays fastest along the direction $e^{-\ii\pi/4}$. The spectral weight has two rows of poles at $k = \pm\nu \pm \ii(n + \frac{1}{2})$, with integer $n \geq 0$, where $\cosh(2\pi k) + \cosh(2\pi\nu)$ vanishes. A rotation of the full real axis would cross some of these poles, so we integrate along the real segment $[0, \nu]$ followed by the ray $\nu + e^{-\ii\pi/4}[0, \infty)$, which passes between them,
    \begin{align}
        G[x_1, x_0; T] = \sum_{j = \pm} \bigg[&\int_0^{\nu} \phi^{j}_k(x_1)\, \phi^{j}_k(x_0)^*\, e^{-\ii\hbar k^2 T/2}\, \dd k 
        + e^{-\ii\pi/4} \int_0^\infty \phi^{j}_{k}(x_1)\, \phi^{j}_{k}(x_0)^*\, e^{-\ii\hbar k^2 T/2}\Big|_{k = \nu + e^{-\ii\pi/4} s}\, \dd s\bigg]
    \end{align}
    for $T > 0$. Along the ray, the integrand decays as the Gaussian $e^{-\hbar s^2 T/2}$, so that we truncate the integral where $e^{-\hbar s^2 T/2} < e^{-80}$ and evaluate both pieces with adaptive Gauss-Kronrod quadrature, to a relative accuracy of about $10^{-15}$. For $|x_0|, |x_1| \lesssim 10$, the hypergeometric functions converge without difficulty. For larger $|x|$, their argument approaches $0$ or $1$ to machine precision, and we replace them by their asymptotic forms \cite{Feldbrugge:2025RosenMorse}. All examples in this paper have $x_0 = -5$. We validate the exact propagator with two independent methods, the spectral representation and the stitching method for real-time path integrals \cite{Feldbrugge:2025Efficient}.

    For the complex Planck constant $1/\hbar = e^{\ii\theta}/|\hbar|$ of \cref{fig:rotation}, the expression continues analytically, with a complex parameter $\nu$. The factor $e^{-\ii\hbar k^2 T/2}$ now decays fastest along the direction $e^{\ii(\theta/2 - \pi/4)}$. We integrate along the ray $e^{\ii(\theta/2 - \pi/4)}[0, \infty)$ from the origin, and subtract $2\pi\ii$ times the residues of the poles $\pm\nu + \ii(n + \frac{1}{2})$ that lie between this ray and the contour $0 \to \nu \to \infty$ that continues the integral at real $\hbar$; the spectral weight has the residue $\tanh(\pi p)/4\pi$ at the pole $p$. Near $\theta = \pi/2$, the contributions $j = \pm 1$ to the residues cancel by many orders of magnitude, and we evaluate them in arithmetic with $256$ bits. A Crank-Nicolson solution of the Schr\"odinger equation confirms the result.

\end{document}